\documentclass[aps,prx,reprint,preprintnumbers,superscriptaddress,nofootinbib,longbibliography,floatfix]{revtex4-1}
\pdfoutput=1
\usepackage{rotating}
\usepackage{array}
\usepackage{amsmath}
\usepackage[normalem]{ulem}
\usepackage{slashed}
\usepackage{booktabs}
\usepackage[pdftex,table]{xcolor}
\usepackage{units}
\usepackage{xfrac}
\usepackage{mathtools}
\usepackage{empheq}
\usepackage[]{units}
\usepackage{multirow}
\usepackage{amssymb}
\usepackage{url}
\usepackage{comment}
\usepackage{physics}
\usepackage{color,soul}
\usepackage{bbm}
\usepackage[caption=false]{subfig}

\usepackage{hyperref}
\hypersetup{
  colorlinks=true,
  citecolor=blue,
  linkcolor=blue,
  urlcolor=blue
}

\newcommand{\A}{\mathbb{A}}
\newcommand{\R}{\mathbb{R}}
\newcommand{\Z}{\mathbb{Z}}

\renewcommand{\O}{\mathcal{O}}
\newcommand{\U}{\mathcal{U}}
\newcommand{\relu}{\operatorname{ReLU}}

\usepackage{color}
\definecolor{darkred}{rgb}{1.0,0.1,0.1}
\definecolor{darkgreen}{rgb}{0.1,0.7,0.1}
\definecolor{darkblue}{rgb}{0.1,0.1,1.0}

\DeclareRobustCommand{\Sec}[1]{Sec.~\ref{sec:#1}}
\DeclareRobustCommand{\Secs}[2]{Secs.~\ref{sec:#1} and \ref{sec:#2}}
\DeclareRobustCommand{\App}[1]{App.~\ref{app:#1}}

\DeclareRobustCommand{\Fig}[1]{Fig.~\ref{fig:#1}}

\DeclareRobustCommand{\Eq}[1]{Eq.~(\ref{eq:#1})}

\DeclareRobustCommand{\Ref}[1]{Ref.~\cite{#1}}

\begin{document}

\title{SymmetryGAN: Symmetry Discovery with Deep Learning}

\author{Krish Desai}
\email{krish.desai@berkeley.edu}
\affiliation{Department of Physics, University of California, Berkeley, CA 94720, USA}
\affiliation{Physics Division, Lawrence Berkeley National Laboratory, Berkeley, CA 94720, USA}

\author{Benjamin Nachman}
\email{bpnachman@lbl.gov}
\affiliation{Physics Division, Lawrence Berkeley National Laboratory, Berkeley, CA 94720, USA}
\affiliation{Berkeley Institute for Data Science, University of California, Berkeley, CA 94720, USA}

\author{Jesse Thaler}
\email{jthaler@mit.edu}
\affiliation{Center for Theoretical Physics, Massachusetts Institute of Technology, Cambridge, MA 02139, USA}
\affiliation{The NSF AI Institute for Artificial Intelligence and Fundamental Interactions}

\preprint{MIT-CTP 5377}

\begin{abstract}

What are the symmetries of a dataset?
Whereas the symmetries of an individual data element can be characterized by its invariance under various transformations, the symmetries of an ensemble of data elements are ambiguous due to Jacobian factors introduced while changing coordinates.
In this paper, we provide a rigorous statistical definition of the symmetries of a dataset, which involves \textit{inertial} reference densities, in analogy to inertial frames in classical mechanics.
We then propose SymmetryGAN as a novel and powerful approach to automatically discover symmetries using a deep learning method based on generative adversarial networks (GANs).
When applied to Gaussian examples, SymmetryGAN shows excellent empirical performance, in agreement with expectations from the analytic loss landscape.
SymmetryGAN is then applied to simulated dijet events from the Large Hadron Collider (LHC) to demonstrate the potential utility of this method in high energy collider physics applications.
Going beyond symmetry discovery, we consider procedures to infer the underlying symmetry group from empirical data.
\end{abstract}

\maketitle

{\small
\tableofcontents
}

\section{Introduction}
\label{sec:intro}

The properties and dynamics of physical systems are closely tied to their symmetries.
Often these symmetries are known from fundamental principles. There are also, however, systems with unknown or emergent symmetries.
Discovering and characterizing these symmetries is an essential component of physics research.

Beyond their inherent interest, symmetries are also practically useful for increasing the statistical power of datasets for various analysis goals.
For example, a dataset can be augmented with pseudodata generated by applying symmetry transformations to existing data, thereby creating a larger training sample for machine learning tasks.
Neural network architectures can be constructed to respect symmetries (e.g.~convolutional neural networks and translation symmetries~\cite{6795724}), in order to improve generalization and reduce the number of model parameters.
Furthermore, symmetries can significantly increase the size of a useful synthetic dataset created from a generative model trained on a limited set of examples~\cite{2008.06545,Dillon:2021gag}.

Deep learning is a powerful tool for identifying patterns in high-dimensional data and is therefore a promising technique for symmetry discovery.
A variety of deep learning methods have been proposed for symmetry discovery and related tasks.
Neural networks can parametrize the equations of motion for physical systems, which can have conserved quantities resulting from symmetries~\cite{greydanus2019hamiltonian,cranmer2020lagrangian}.
Generic neural networks targeting classification tasks can encode symmetries in their hidden layers~\cite{Barenboim:2021vzh,Krippendorf:2020gny}. %
This possibility can be used to actively learn symmetries by encoding a shared equivariance in hidden layers across learning tasks~\cite{zhou2021metalearning}.
Directly learning symmetries can be framed as an inference problem given access to parametric symmetry transformations of the same dataset~\cite{benton2020learning}.
A given symmetry can be identified in data if a classifier is unable to distinguish a dataset from its symmetric counterpart~\cite{Tombs:2021wae,Lester:2021kur,Lester:2021aks} (similar to anomaly detection methods comparing data to a reference~\cite{Collins:2018epr,Collins:2019jip,DAgnolo:2018cun}).
Another class of targeted approaches can be found in the domain of automatic data augmentation.
If a dataset can be augmented without changing its statistical properties, then one has learned a symmetry. Significant advances in this area have used reinforcement learning~\cite{cubuk2019autoaugment,lim2019fast}.

\begin{figure}
    \centering
    \includegraphics[width=0.45\textwidth]{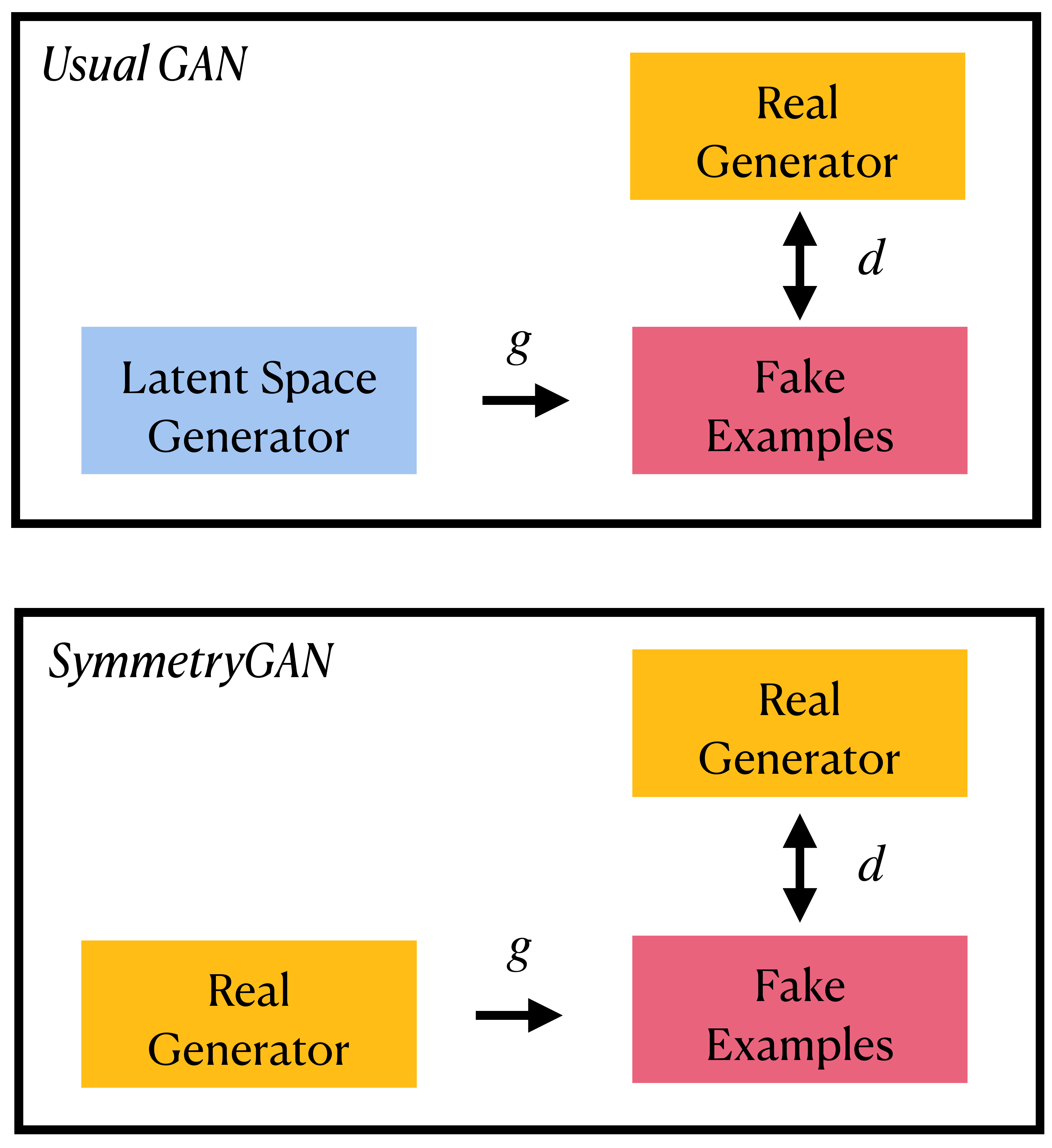}
    \caption{
    A schematic diagram of (top) the training setup for a usual GAN and (bottom) the SymmetryGAN variation discussed in this paper for automatically discovering symmetries.
    Here, $g$ is the generator and $d$ is the discriminator.
    Not represented here is the incorporation of the inertial reference dataset.
    In our numerical examples, this is accomplished by directly imposing constraints on $g$.
    }
    \label{fig:schematic}
\end{figure}

An alternative symmetry discovery approach that is flexible, fully differentiable, and simple is based on generative models~\cite{hataya2019faster,antoniou2018data}.
Usually, a generative model is a function that maps random numbers to structured data.
For example, a deep generative surrogate model can be trained such that the resulting probability density matches that of a target dataset.
For symmetry discovery, by contrast, the random numbers are replaced with the target dataset itself.
In this way, a well-trained generator designed to confound an adversary will implement a symmetry transformation.  
We call this generative model framework for symmetry discovery \emph{SymmetryGAN}, since it has the same basic training strategy as a generative adversarial network (GAN)~\cite{Goodfellow:2014upx}, as shown in \Fig{schematic}.

In this paper, we extend the SymmetryGAN approach (suggested in \Ref{lim2019fast}, but in the language of data augmentation rather than symmetries) and introduce it to the physics community.
In particular, we build a rigorous statistical framework for describing the symmetries of a dataset and construct a learning paradigm for automatically detecting generic symmetries.
The key idea is that symmetries of a target dataset have to be defined with respect to an \emph{inertial} reference dataset, analogous to inertial frames in classical mechanics.
Our deep learning setup is simpler than existing approaches and we develop an analytic understanding of the algorithm's performance in simple cases.
This in turn allows us to understand the dynamics of the machine learning as it trains from a random initialization to an element of the symmetry group.
The primary purpose of this paper is to carefully demonstrate that this method of symmetry discovery works.
Having done so, in the last section we move forward to a discussion of methods to infer which formal groups of symmetries are present in the dataset, a related but distinct problem which is a rich area for future research.

 This rest of this paper is organized as follows.
 In \Sec{stats}, we build a rigorous statistical framework for discovering the symmetries of a dataset, contrasting it with discovering the symmetries of an individual data element.
 Our machine learning approach with an inertial restriction is introduced in \Sec{MLa} and the deep learning implementation is described in \Sec{ML}.
 Empirical studies of simple Gaussian examples, including both analytic and numerical results, are presented in \Sec{results}.
 We then apply our method to a high energy physics dataset in \Sec{hepexample}.
  In \Sec{inference}, we discuss possible ways to go beyond symmetry discovery and towards symmetry inference, with further studies in \App{symmetry_discovery_map}.
 Our conclusions and outlook are in \Sec{conclusions}.

\section{Statistics of Symmetries}
\label{sec:stats}

What is a symmetry?
Let $X$ be a random variable on an open set $\O\subseteq\R^n$, and let $x$ be an instantiation of $X$.
When we refer to the symmetry of an individual data element $x \in X$, we usually mean a transformation $h:\O\rightarrow\O$ such that:
\begin{equation}
    h(x) = x,
\end{equation}
i.e.\ $x$ is invariant to the transformation $h$.
More generally, we can consider functions of individual data elements, $f:\O\subseteq\mathbb{R}^n\rightarrow\O'\subseteq\mathbb{R}^m$.
In that case, the function is symmetric if
\begin{equation}
\label{eq:functionsymmetry}
    f(h(x)) = f(x),
\end{equation}
i.e. the output of $f$ is invariant to the transformation $h$ acting on $x$.
One can also consider equivariances, where the output of $f$ has well-defined transformation properties under the symmetry~\cite{Dolan:2020qkr,serviansky2020set2graph,Bogatskiy:2020tje,Shimmin:2021pkm}.
While symmetries acting on individual data elements are interesting, they are \emph{not} the focus of this paper.

We are interested in the symmetries of a dataset as a whole, treated as a statistical distribution.
Let $X$ be governed by the probability density function (PDF) $p$.
Naively, a symmetry of the dataset $X$ is a map $g:\mathbb{R}^n\rightarrow\mathbb{R}^n$ such that $g$ preserves the PDF:
\begin{equation}
\label{eq:naivesymmetry}
p(X = x) = p(X = g(x)) \, |g'(x)|,
\end{equation}
where $|g'(x)|$ is the Jacobian determinant of $g$.
While it is necessary that any candidate symmetry preserves the probability density, it is not sufficient, at least not in the usual way that we, as particle physicists, think about symmetries.

Consider the simple case of $n=1$.
Let $F$ be the cumulative distribution function (CDF) of $X$.
$F(X)$ is itself a random variable satisfying 
\begin{equation}
    F(X)\sim\mathcal{U}[0,1],
\end{equation}
where $\mathcal{U}(\O)$ is the uniform random variable on $\O$.
Conversely, $F^{-1}(\mathcal{U}[0,1])$ is a random variable governed by the PDF $p$ (for technical details, see~\Ref{10.2307/2132726}).
The uniform distribution on the interval $[0,1]$ has many PDF-preserving maps, such as the quantile inversion map:
\begin{equation}
\widetilde{g}(x)=1-x.
\end{equation}
This map has the additional property that $\widetilde{g}(\widetilde{g}(x))=x$, so it appears to represent a $\mathbb{Z}_2$ (i.e.~parity) symmetry.
Using the CDF map from above, every probability density $p$ admits a $\mathbb{Z}_2$ PDF-preserving map:
\begin{equation}
\label{eq:PDFpreserveZ2}
    g=F^{-1}\circ\widetilde{g}\circ F,
\end{equation}
where $\circ$ refers to functional composition.

If we were to accept \Eq{naivesymmetry} as the definition of a symmetry, then \emph{all} one-dimensional random variables would have a $\mathbb{Z}_2$ symmetry, namely the one in \Eq{PDFpreserveZ2}.
While true in a technical sense, this is not what particle physicists (or, to our knowledge, any domain experts) think of as a symmetry of a dataset.
The precise definition of a symmetry must therefore be stricter than simply PDF-preserving.
In particular, while this $\mathbb{Z}_2$ PDF-preserving map applies to every one-dimensional random variable, it requires a different map for each such variable.
When we usually think about symmetries, we imagine common maps that can be applied to a variety of physical systems that share the same underlying symmetry structure.

This line of thinking suggests a sharper definition of a symmetry that makes use of a reference distribution.
Consider two probability density functions
\begin{equation}
\label{eq:twoPDFs}
p:\mathbb{R}^n\rightarrow\mathbb{R}_{\ge0}, \quad p_I:\mathbb{R}^n\rightarrow\mathbb{R}_{\ge0},
\end{equation}
where $\mathbb{R}_{\ge0}$ is the set of non-negative real numbers.
A map $g:\mathbb{R}^n\rightarrow\mathbb{R}^n$ is defined to be a symmetry of $p$ \emph{relative} to $p_I$ if it is PDF-preserving for both $p$ and $p_I$:
\begin{equation}
\label{eq:improvedsymmetry}
p(x) = p(g(x)) \, |g'(x)|, \quad p_I(x) = p_I(g(x)) \, |g'(x)|.
\end{equation}
The reference or \textit{inertial} density $p_I$ is the analogue of an inertial reference frame in classical mechanics.
This new definition of a symmetry will typically exclude quantile maps, like $\widetilde{g}$ above, because the $\widetilde{g}$ that works for one random variable will typically not work for another (e.g.\ Gaussian and exponential random variables).

While this new definition solves the problem of ``fake'' symmetries, it also introduces a dependence on the inertial distribution.  
Just as with inertial reference frames, however, there is often a canonical choice for $p_I$ which reduces the number of possibilities in practice.
A natural choice for many physics datasets is to pick the uniform distribution on $\mathbb{R}^{n}$, where $n$ is the dimension of the dataset, because many physics problems outside of General Relativity are set either in Euclidian space $\R^n$ or in Lorentzian space $\R^{p, q}$, and the affine groups discussed above are independent of signature $\operatorname{Aff}_{p, q}(\R) = \operatorname{Aff}_{p+q}(\R)$.
This not a proper (i.e.\ normalizable) probability density because $R^n$ is a non-compact space,\footnote{A compact space is a topological space that is closed and bounded.} so we discuss techniques below to use it as the inertial distribution nonetheless.

Finally, it is instructive to relate the definitions of symmetries for datasets and functions.
Given the two PDFs in \Eq{twoPDFs}, we can construct the likelihood ratio
\begin{equation}
    \ell(x) \equiv \frac{p(x)}{p_I(x)}.
\end{equation}
Applying the symmetry map $g$ as in \Eq{improvedsymmetry}, the likelihood ratio transforms as:
\begin{equation}
    \ell(g(x)) = \frac{p(g(x))}{p_I(g(x))} = \frac{p(x)}{p_I(x)} = \ell(x),
\end{equation}
where the Jacobian factor $|g'(x)|$ cancels between the numerator and denominator.
Therefore the likelihood ratio, which is an ordinary function, is symmetric by the definition in \Eq{functionsymmetry}.
This cancelling of the Jacobian factor is an intuitive way to understand why an inertial reference density is necessary to define the symmetry of a dataset.

\section{Machine Learning with Inertial Restrictions}
\label{sec:MLa}

The SymmetryGAN paradigm for discovering symmetries in a dataset involves simultaneously learning two functions:
\begin{align}
    g:\mathbb{R}^n &\rightarrow\mathbb{R}^n,\\
    d:\mathbb{R}^n &\rightarrow[0,1].
\end{align}
The function $g$ is a \emph{generator} that represents the symmetry map.%
\footnote{Here, we are using the machine learning meaning of a ``generator'', which differs from the generator of a symmetry group, though they are closely related.}
The function $d$ is a \emph{discriminator} that tries to distinguish the input data $\{x_i\}$ from the transformed data $\{g(x_i)\}$.
When the discriminator cannot distinguish the original data from the transformed data, then $g$ will be a symmetry.
The technical details of this approach are provided in \Sec{ML} using the framework of adversarial networks.
The generator is randomly initialized on the search manifold, and through gradient descent, converges to the nearest symmetry.
It is possible that the nearest symmetry is the identity transformation, in which case the generator will converge to the identity map.
When the generator is randomly initialized across the search manifold several times, however, there is no reason why the nearest symmetry on that manifold should always be the identity map, so the generator will converge to the nearest non-trivial symmetry.
In fact, when the dataset respects a continuous symmetry group, the probability of the generator converging to the identity is zero.

As described in \Sec{stats}, it is not sufficient to require that $g$ preserves the PDF of the input data; it also has to preserve the PDF of the inertial density.
There are several methods to implement an inertial restriction into the machine learning strategy.
\begin{itemize}
    \item \emph{Simultaneous discrimination:}
    In this method, the discriminator $d$ is applied both to the input dataset and to data drawn from the inertial density $p_I$.
    The training procedure penalizes any map $g$ that does not fool $d$ for both datasets.
    In practice, it might be advantageous to use two separate discriminators $d$ and $d_I$ for this approach.
    \item \emph{Two stage selection:}
    Here, one first identifies all PDF-preserving maps $g$.
    Then one \textit{post hoc} selects the ones that also preserve the inertial density.
    \item \emph{Upfront restriction:}
    If the PDF-preserving maps of $p_I$ are already known, then one could restrict the set of maps $g$ at the outset.
    This allows one to perform an unconstrained optimization on the restricted search space.
\end{itemize}

Each of these methods has advantages and disadvantages.
The first two options require sampling from the inertial density $p_I$.
This is advantageous in cases where the symmetries of the inertial density are not known analytically.
When $p_I$ is uniform on $\mathbb{R}^n$ or another unbounded domain, though, these approaches are not feasible.%
\footnote{One could try to leverage approximate strategies, such as cutting off the support for $p_I$ a few standard deviations away from the mean of $p$.  Still, one can run into edge effects if there is a mismatch between the domain and range of $g$.}
The second option is computationally wasteful, as the space of PDF-preserving maps is generally much larger than the space of symmetry maps.
We focus on the third option: restricting the set of functions $g$ to be automatically PDF-preserving for $p_I$.
This in turn requires a way to parametrize all such $g$, or at least a large subset of them.

For all of the studies in this paper, we further focus on the case where the inertial distribution $p_I$ is uniform on $\mathbb{R}^n$.
For any open set $\O\subseteq\R^n$, a differentiable function $g:\O\to \O$ preserves the PDF of the uniform distribution $\U(\O)$ if and only if $g$ is an equiareal map.%
\footnote{By carefully taking suitable limits, these ideas go through even if $\U(\O)$ is an improper prior. The important takeaway is that uniform distributions are preserved by equiareal maps.}
To see this, note that the PDF of $X\sim\U(\O)$ is $p(X = x) = 1/\operatorname{Vol}(\O)$, where $\operatorname{Vol}$ is the $n-$volume.
Hence, the PDF-preserving condition $p = p\circ g\cdot |g'|$ is met if and only if $|g'| = 1$.
A map is equiareal if and only if its Jacobian determinant is $1$, which proves our claim.
Therefore, our search space to discover symmetries of physics datasets will be the space of equiareal maps of appropriate dimension.
Of course, there are interesting physics symmetries that do not preserve uniform distributions on $\mathbb{R}^n$; these would require an alternative approach.

The set of equiareal maps for $n > 1$ is not well characterized.
For example, even for $n = 2$, not all equiareal maps are linear.
A simple example of a non-linear area-preserving map is the H\'{e}non map~\cite{10.2307/43635985}: $g(x,y)=(x,y-x^2)$.
This makes the space of equiareal maps difficult to directly encode into the learning.
While the general set of equiareal maps is difficult to parametrize, the set of area preserving linear maps on $\mathbb R^n$ is well understood:
\begin{multline*}
    \A SL^\pm_n(\mathbb{R}) = \{g: \R^n\to\R^n \; | \; g(x) = Mx + V,\\ M\in \R^{n\times n}, \det M = \pm 1, V\in \R^n\}.
\end{multline*}
This is a subgroup of the general affine group $\operatorname{Aff}_n(\R)$, and it can be characterized as a topological group\footnote{A topological group is a topological space with a group operation defined on it, and where the group operation and inversion are continuous functions.} of dimension $n(n+1) - 1$.
These maps even have complete parametrizations such as the \textit{Iwasawa decomposition}~\cite{10.2307/1969548} which significantly aid the symmetry discovery process.

Not all symmetries are linear, however, and if one chooses $\A SL^\pm_n(\mathbb{R})$ as the search space, one cannot discover non-linear maps.
Even so, the subset of symmetries discoverable within $\A SL^\pm_n(\mathbb{R})$ is rich enough, and the benefits of having a known parametrization valuable enough, that we focus on linear symmetries in this paper and leave the study of non-linear symmetries to future work.

\section{Deep Learning Implementation}
\label{sec:ML}

To implement the SymmetryGAN procedure, we modify the learning setup of a GAN~\cite{Goodfellow:2014upx}.
For a typical GAN, a generator function $g$ surjects a latent space onto a data space.%
\footnote{While all the GANs discussed here are (approximately) bijective, GANs in general need not be. Symmetry discovery requires the generator to be bijective, so one may want to leverage nomalizing flows~\cite{10.5555/3045118.3045281,Kobyzev2020} in future work.}
Then, a discriminator distinguishes generated examples from target examples.

For a SymmetryGAN, the latent probability density is the \emph{same} as the target probability density, as illustrated in \Fig{schematic}.
The generator $g$ and discriminator $d$ are parametrized as neural networks.
Following \Sec{MLa}, we construct the generator $g$ such that it is guaranteed to preserve the inertial distribution, e.g.\ it is an area-preserving linear transformation, but the discriminator $d$ has no such restriction.
These two neural networks are then trained simultaneously to optimize the binary cross entropy loss functional, where the generator tries to maximize the loss with respect to $g$ and the discriminator tries to minimize the loss with respect to $d$.
The binary cross entropy loss functional is:
\begin{align}
\label{eq:numericloss}
    L[g,d]=-\frac1N\sum_{x\in\{x_i\}_{i=1}^N}\Big[\log\big(d(x)\big) + \log\big(1-d(g(x))\big)\Big]\,.
\end{align}
This differs from the usual binary cross entropy in that the same samples appear in the first and second terms.
A similar structure appears in neural resampling~\cite{Nachman:2020fff} and in step 2 of the \textsc{OmniFold} algorithm~\cite{Andreassen:2019cjw}.

We now show that optimizing the above loss corresponds to finding a symmetry generator $g$.
The behavior of \Eq{numericloss} can be understood analytically by considering the limit of infinite data:
\begin{align}\nonumber
    L[g,d]&=-\int \Big[\log\big(d(x)\big) \, p(x)\\\label{eq:loss}
    &\qquad+\log\big(1-d(g(x))\big)\, p(g(x))\, |g'(x)|\Big]\dd x\,,
\end{align}
where the Jacobian factor $|g'(x)|$ is now made manifest.
For a fixed $g$, the optimal $d$ is the usual result from binary classification (see e.g.\ \Ref{hastie01statisticallearning,sugiyama_suzuki_kanamori_2012}):
\begin{align}
\label{eq:optimalf}
    d_*=\frac{p(x)}{p(x)+p(g(x)) \, |g'(x)|}\,,
\end{align}
which is the ratio of the probability density of the first term in \Eq{loss} to the sum of the densities of both terms.
We then insert $d_*$ into \Eq{loss} and optimize using the Euler-Lagrange equation:
\begin{equation}
\frac{\delta L[g,g']}{\delta g}=\frac{\partial L}{\partial g}-\frac{\dd}{\dd x} \frac{\partial L}{\partial g'} = 0.
\end{equation}
By use of a computer algebra system capable of solving simple differential equations (in our case, Mathematica~\cite{Mathematica}), one can show that the optimal $g$ satisfies
\begin{equation}
p(x)=p(g_*(x)) \, |g_{*}'(x)|,
\end{equation}
i.e.\ $g$ is PDF preserving as in \Eq{naivesymmetry}.
For such a $g$, we have that $d_* = \frac12$, the loss is maximized at a value of $2 \log 2$, and the discriminator is maximally confounded.

The SymmetryGAN approach has the potential to find any symmetry representable by $g(x)$.
To target a particular symmetry subgroup, $G \leq \A SL_n^\pm(\R)$, we can add a term to the loss function.
For example, to discover a cyclic symmetry group, $G = \Z_q, q\in\mathbb N$, the loss function can be augmented with a mean squared error term:
\begin{align}
\label{eq:cyclicloss}
    L[g,d] = L_\text{BCE}[g,d]-\frac\alpha N\sum_{x\in\{x_i\}_{i=1}^N}(g^q(x) - x)^2,
\end{align}
where $L_\text{BCE}$ is the binary cross entropy loss in \Eq{numericloss}, $g^q$ is $g$ composed with itself $q$ times, and $\alpha>0$ is a weighting hyperparameter.
A SymmetryGAN with this loss function will discover the largest subgroup of $G$ that is a symmetry of the dataset.

\section{Empirical Gaussian Experiments}
\label{sec:results}

In this section, we study the SymmetryGAN approach both analytically and numerically in a variety of simple Gaussian examples.
For the empirical studies here and in \Sec{hepexample}, all neural networks are implemented using the \textsc{Keras}~\cite{keras} high-level API with the \textsc{Tensorflow2} backend~\cite{tensorflow} and optimized with \textsc{Adam}~\cite{adam}.
The generator function $g$ is parametrized as a linear function, with constraints that vary by example and are described further below.
The discriminator function $d$ is parametrized with two hidden layers, using 25 nodes per layer.
Rectified Linear Unit (ReLU) activation functions are used for the intermediate layers and a sigmoid function is used for the last layer.
For the empirical studies, $128$ events are generated for each example.

\subsection{One-Dimensional Gaussian}
\label{sec:1d_example}

Our first example involves data drawn from a one-dimensional Gaussian distribution with a $\mathbb{Z}_2$ reflection symmetry.
Data are distributed according to the probability distribution $\mathcal N(0.5, 1.0),$ i.e.\ a Gaussian with $\mu = 0.5$ and $\sigma^2 = 1.0$.
This distribution has precisely two symmetries, both linear:
\begin{equation}
\label{eq:1D_minima}
g(x) = x, \qquad g(x) = 1-x.
\end{equation}

\begin{figure}[t]
    \centering
    \includegraphics[width=0.45\textwidth]{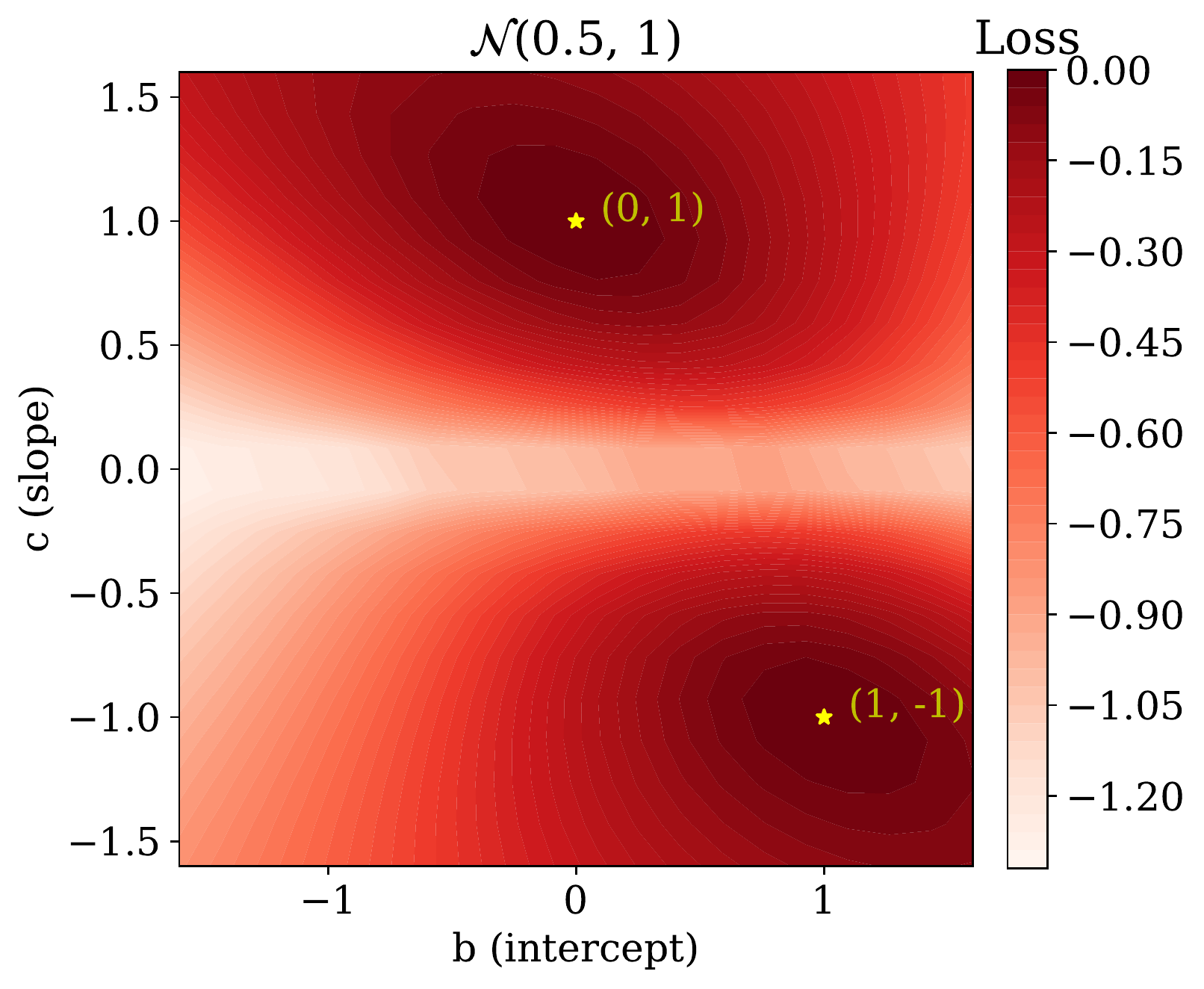}
    \caption{
    The analytic loss landscape in the slope ($c$) vs.\ intercept ($b$) space for the one-dimensional Gaussian example.
    The two maxima are indicated by stars.}
    \label{fig:Z2analytic}
\end{figure}

Implicitly, we are taking the inertial distribution to be uniform on $\mathbb{R}$.
As stated earlier, the PDF-preserving maps of $\U(\R)$ are equireal.
In one dimension, the only equireal maps are linear.
Linear maps in one dimension are defined by two numbers, so the generator function can be parametrized as
\begin{equation}
    \label{eq:linear_form}
    g(x) = b + c \, x. 
\end{equation}
In \Fig{Z2analytic}, we show the analytically computed loss from \Eq{loss} as a function of $b$ and $c$.
In this figure, the discriminator $d$ is taken to be the analytic optimum in \Eq{optimalf}.
There are two maxima in the loss landscape, one corresponding to each of the linear symmetries from \Eq{1D_minima}.
Here, and in most subsequent examples below, we have shifted the output such that maximum loss value is $0$.

Another interesting feature of the loss landscape is the deep minimum at $c=0$ that divides the space into two parts.
This gives rise to the prediction that, under gradient descent, the neural network will find $g(x)= 1-x$ when $c$ is initialized negative and find $g(x) = x$ when $c$ is initialized positive.
In the edge case when $c$ is initialized to precisely zero, the generator is degenerate and no longer even bijective and the outcome is indeterminate, but the likelihood of sampling $c$ to be precisely zero is, of course, zero.
For the rest of the paper, we ignore such edge cases.
There are no such features in the loss landscape as a function of  $b$, suggesting that there should be little dependence on the initial value of $b$.

\begin{figure*}[t]
    \centering
    \subfloat[]{\includegraphics[width=0.31\textwidth]{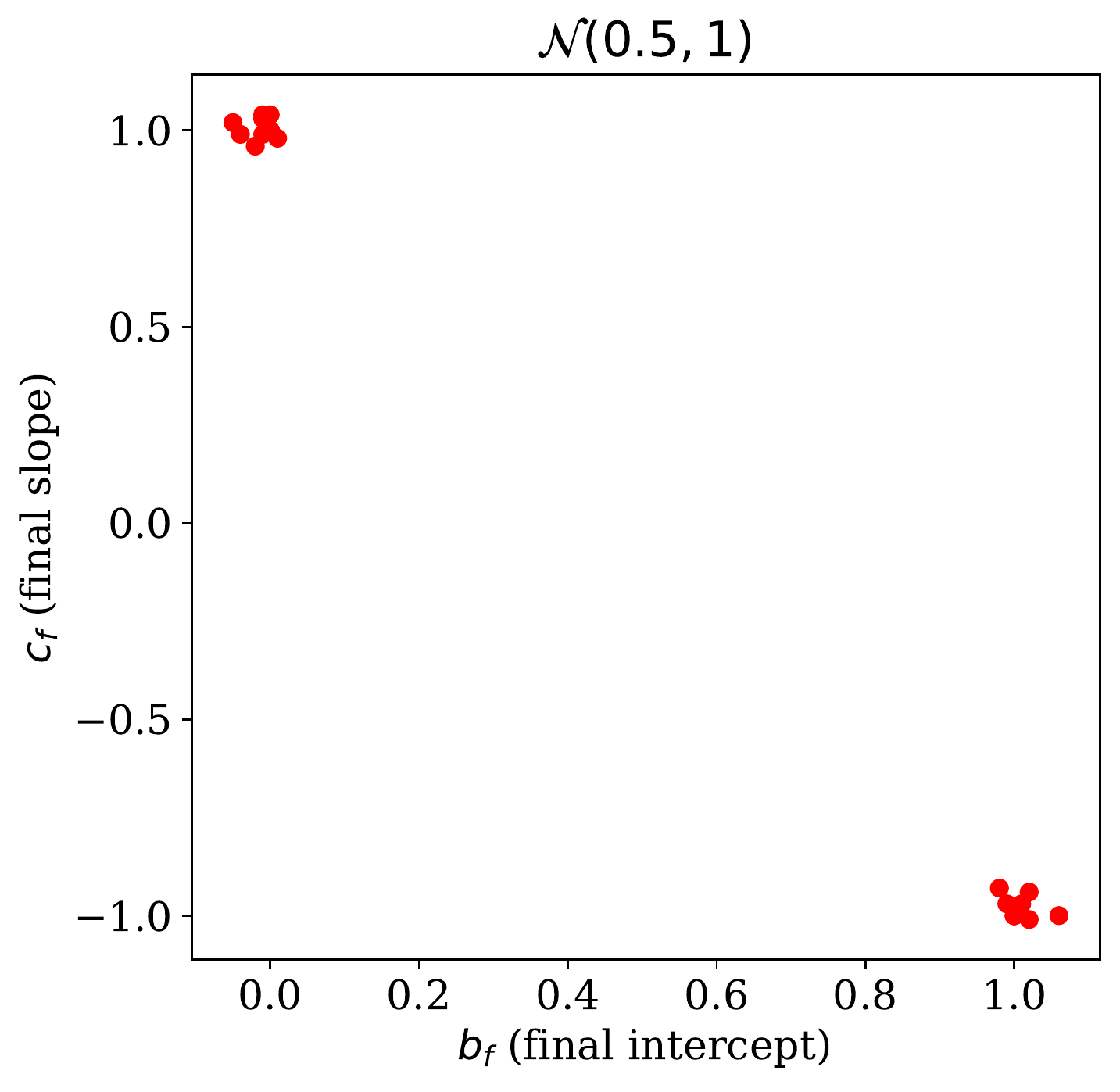}
    \label{fig:Z2numeric_i}}
    $\quad$
    \subfloat[]{\includegraphics[width=0.31\textwidth]{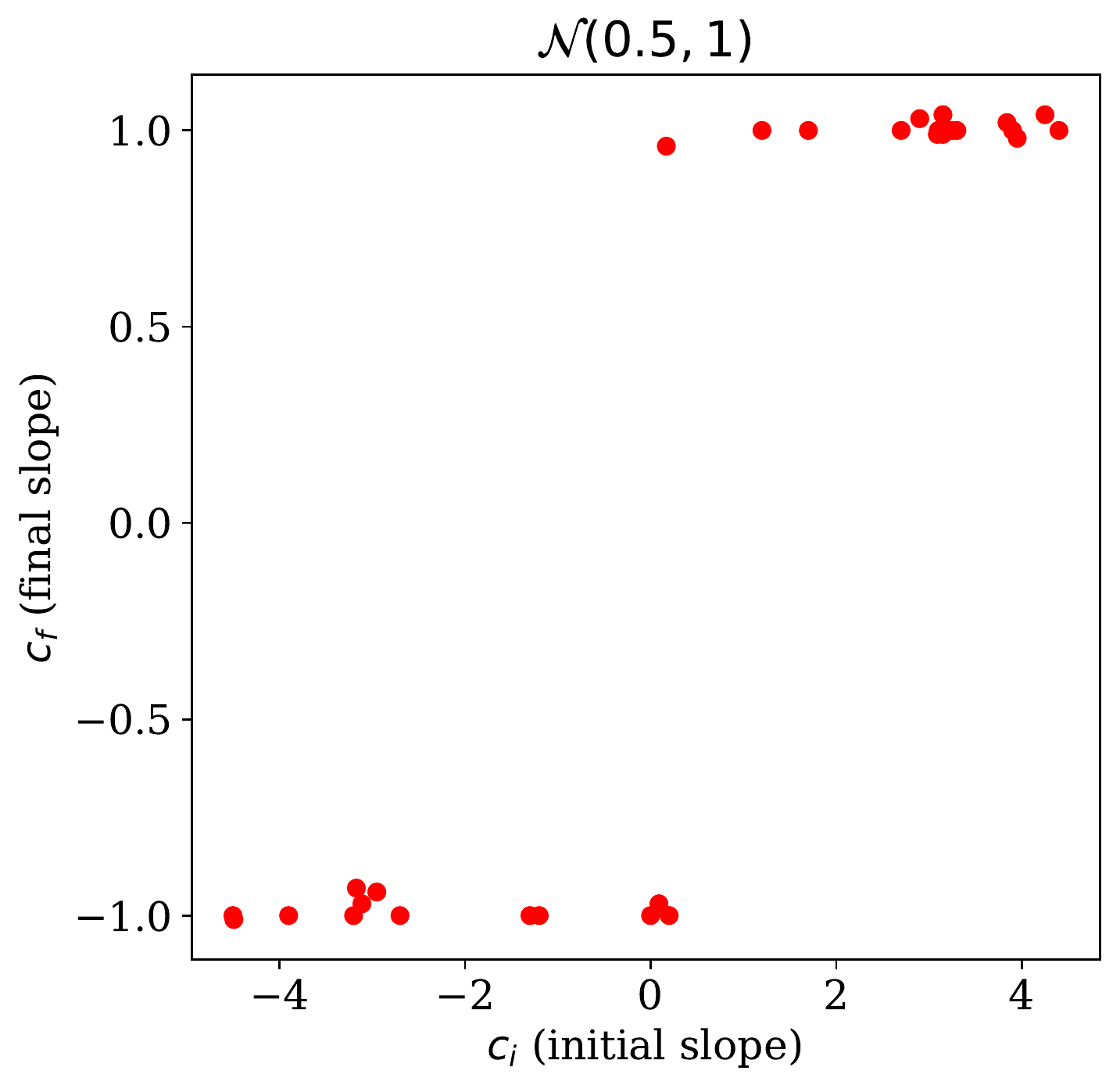}
    \label{fig:Z2numeric_ii}}
    $\quad$
    \subfloat[]{\includegraphics[width=0.31\textwidth]{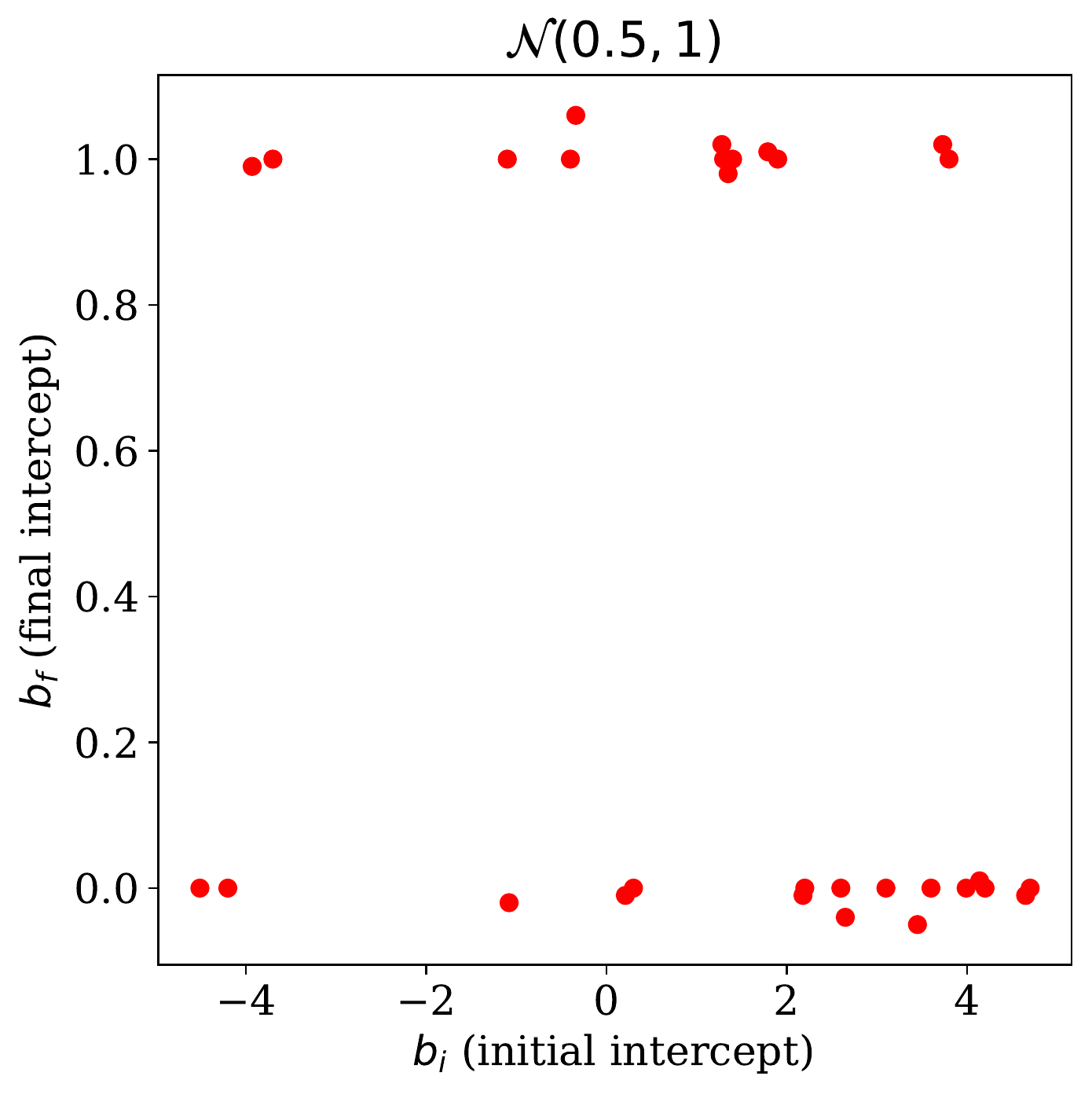}
    \label{fig:Z2numeric_iii}}
    \caption{
    The empirical symmetry discovery process for the one-dimensional Gaussian example.
    The initial parameters have a subscript $i$ and the final parameters have a subscript $f$.
    (i) Final slope ($c_f)$ vs.\ final intercept ($b_f$), showing that the network finds the two maxima.
    (ii) Final slope ($c_f)$ vs.\ initial slope ($c_i$), showing the phase transition at $c_i = 0$.
    (iii) Final intercept ($b_f)$ vs.\ initial intercept ($b_i$), showing the independence on $b_i$.}
    \label{fig:Z2numeric}
\end{figure*}

    These predictions are tested empirically in \Fig{Z2numeric}, where the initialized parameters are $(b_i,c_i)\sim \mathcal{U}([-5, 5]^2)$ and the learned parameters are $(b_f,c_f)$.
    In \Fig{Z2numeric_i}, there are distinct clusters at $(b_f, c_f) = (0, 1)$ and $(1, -1)$,
    showing that the SymmetryGAN correctly finds both symmetries of the distribution and nothing else.
    In \Fig{Z2numeric_ii}, there is a demonstration of the loss barrier in slope space; if the initial slope is positive, the final slope is $+1$, whereas if the initial slope is negative, the final slope is $-1$.
    Finally, \Fig{Z2numeric_iii} shows the absence of a loss barrier in intercept space; the final intercepts are scattered between $0$ and $1$ independent of the initialized intercept. 
    We discuss further the \emph{symmetry discovery map} from initialized to learned parameters in \Sec{symmetry_discovery_map} and \App{symmetry_discovery_map}.
    
\begin{figure}
    \centering
    \includegraphics[width=0.45\textwidth]{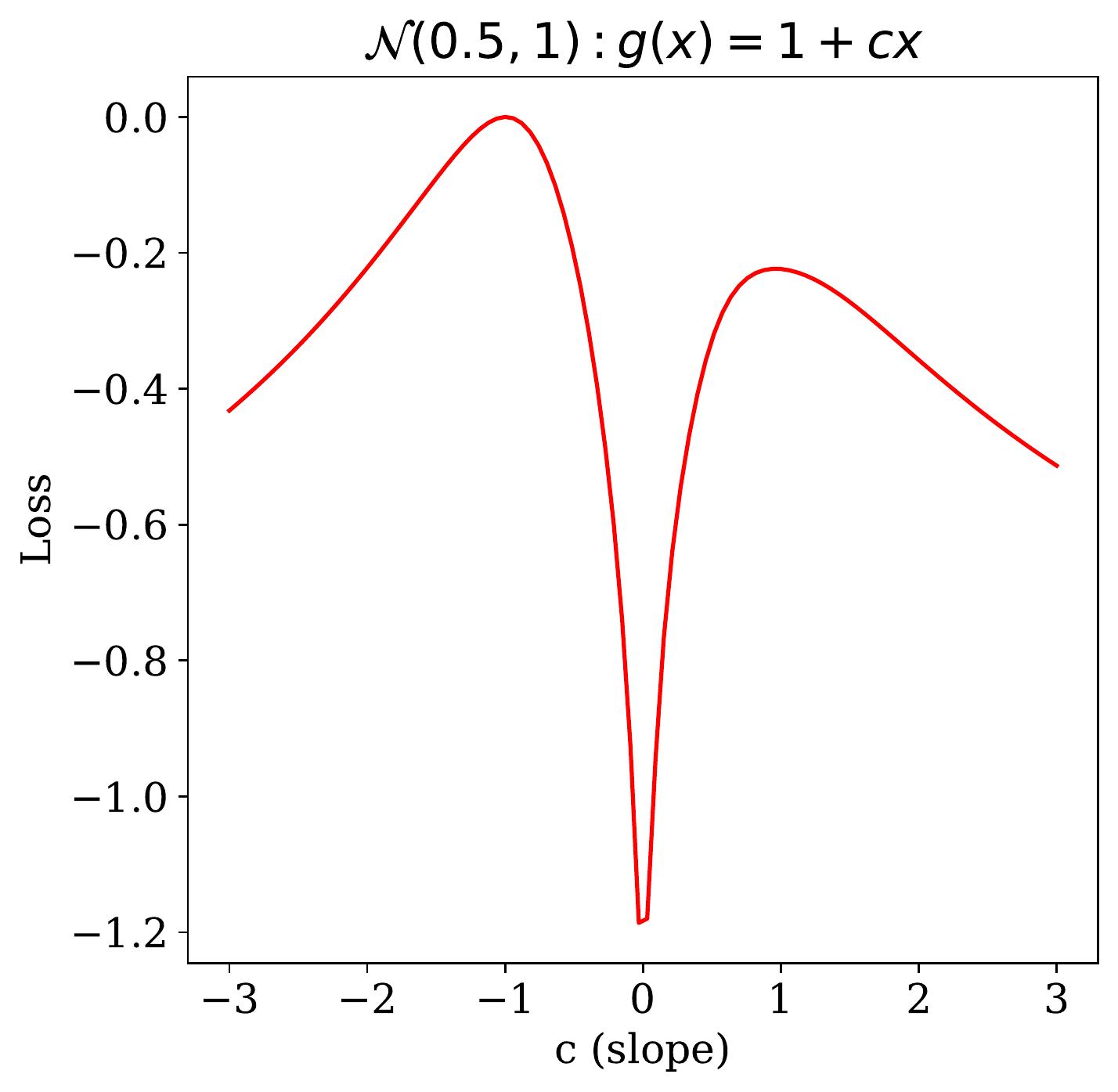}
    \caption{
    The analytic loss landscape for the restricted generator $g(x) = 1 + cx$, with two local maxima at $c = -1$ and $c = 0.5$.}
    \label{fig:Z2restricted}
\end{figure}
    
In the above example, the parametrization of $g$ was sufficiently flexible that the SymmetryGAN could find both symmetries and the loss landscape had no other maxima.
If the space is incompletely parametrized, though, then local maxima can manifest as false symmetries.
For example, suppose instead of a two parameter $g$ as above, $g$ were parameterized as $g(x) = 1 + c\, x$.
The corresponding analytic loss landscape is shown in \Fig{Z2restricted}.
A SymmetryGAN initialized with a negative slope correctly finds the only symmetry of this form, $g(x) = 1 - x$, but a neural network initialized with positive slope is unable to cross over the loss barrier at $c = 0$ and instead settles at the locally loss maximizing $g(x) = 1 + 0.5 \,x$.
While our investigations of $\A SL_n^\pm(\R)$ suggest that this does not happen with the full parametrization, the topology of the set of equiareal maps is not known and therefore obstructions like the one illustrated here are possible.
It is always possible to check if a solution is a symmetry, however.
Specifically, one can apply the learned function to the data and train a \textit{post hoc} discriminator to ensure that its performance is equivalent to random guessing.
For an analytic symmetry, we know that at the point of loss maximization  $p = p\circ g\cdot |g'|$, and consequently $d = \frac{p}{p + p\circ g\cdot |g'|} = \frac12$.
Hence, at the global (symmetry) maxima, $L = - \frac1N\sum_{x_i}\qty[\log d + \log (d\circ g)] = 2\log2$.
On the other hand, there is no way for the neural network to get stuck at non-symmetry local maxima with $L = 2 \log 2$.
Hence, the true symmetries can be distinguished from local optima by checking the value of the loss.

\begin{figure*}[t]
    \centering
    \subfloat[]{\includegraphics[width=0.45\textwidth]{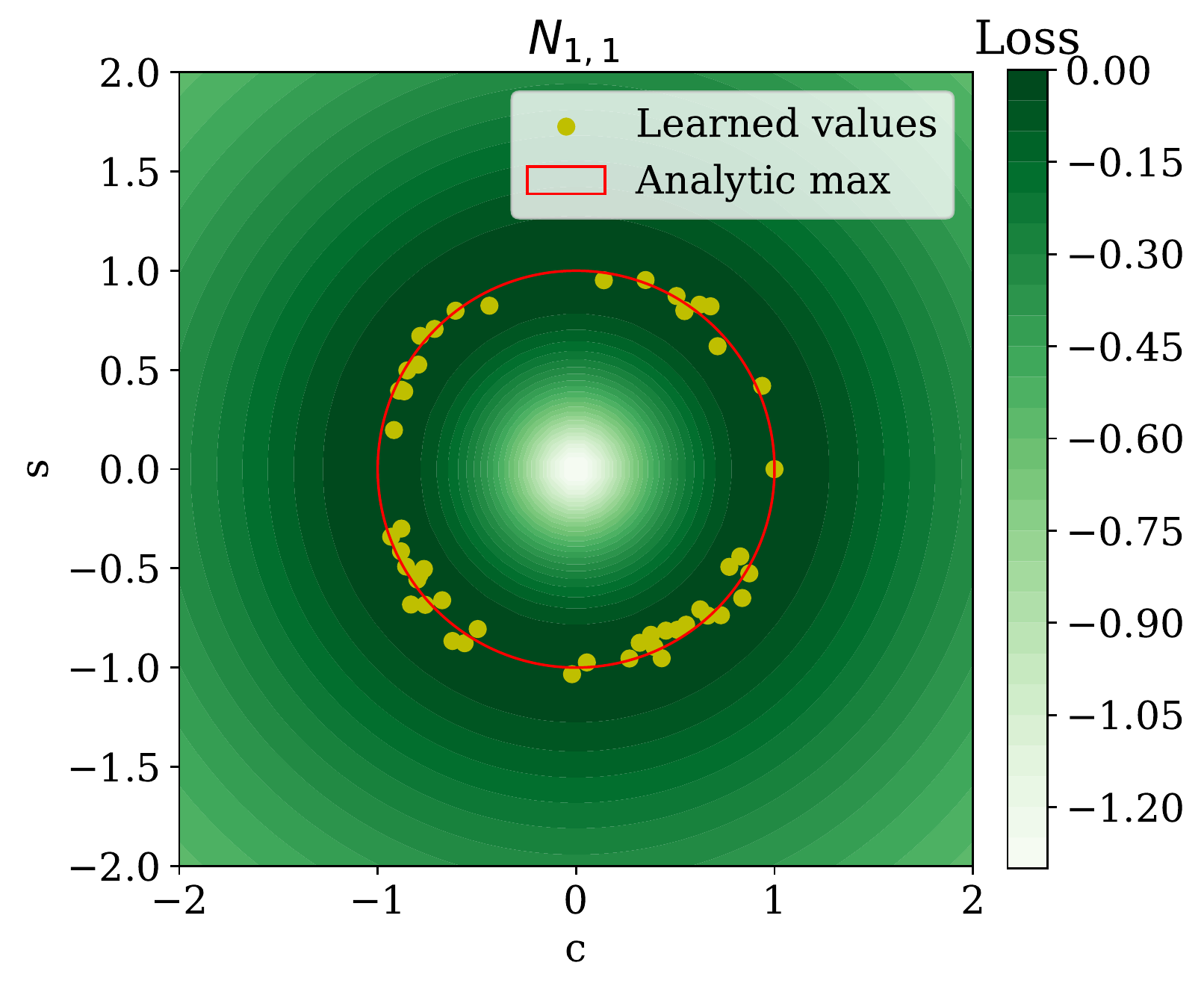} \label{fig:SO2_i}}
    $\quad$
    \subfloat[]{\includegraphics[width=0.45\textwidth]{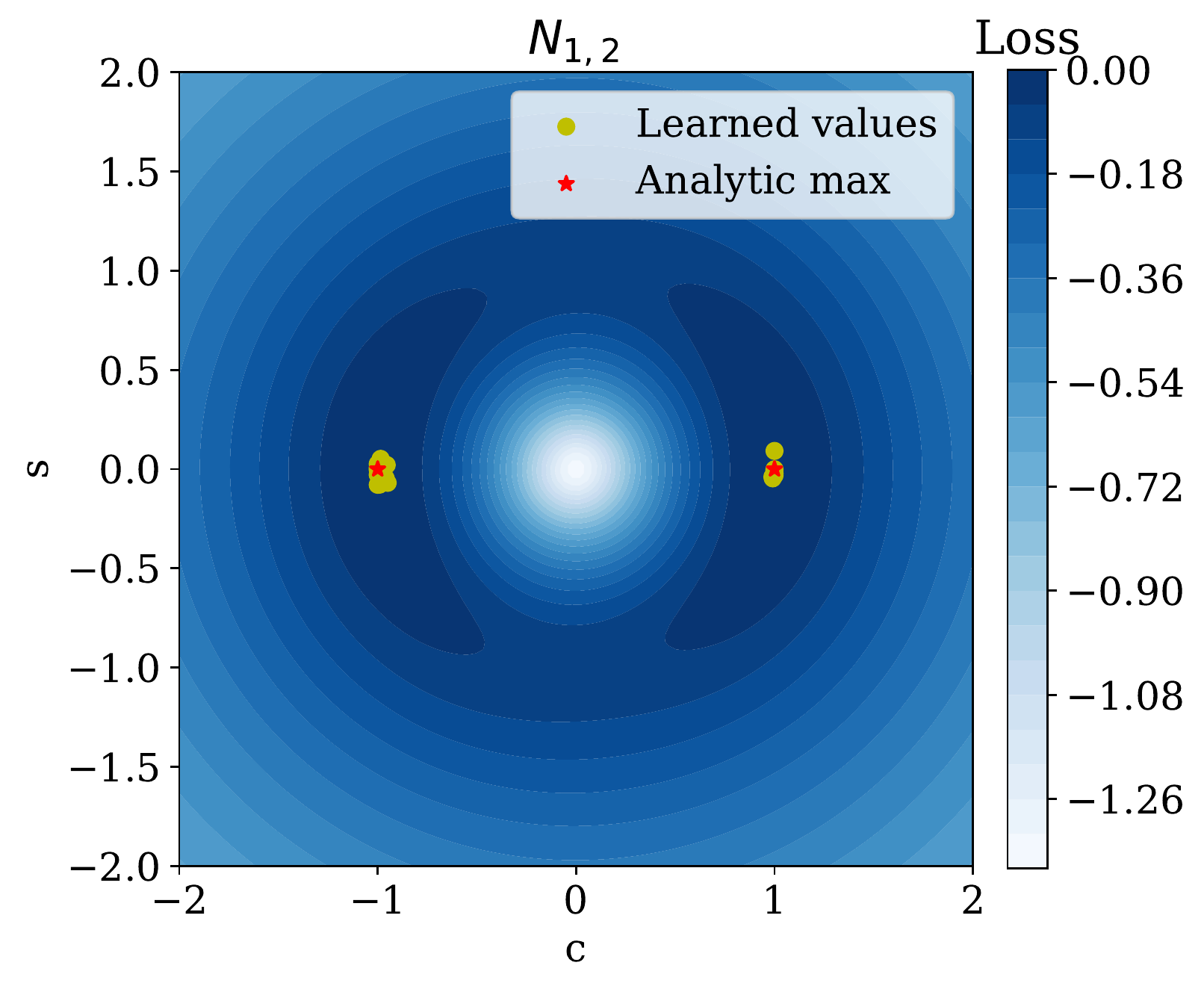}
    \label{fig:SO2_iii}}
    \caption{
    The analytic loss landscapes overlaid with empirically discovered symmetries for the two-dimensional Gaussian examples with the generator restriction in \Eq{rotation}.
    (i) The Gaussian $N_{1,1}$ with uniform covariance, which has loss maxima on the unit circle $c^2 + s^2 = 1$.
(ii) The Gaussian $N_{1,2}$ whose covariance matrix has non-equal diagonal elements, which only has symmetries at $c = \pm 1$ and $s = 0$.
    }
    \label{fig:SO2}
\end{figure*}

\subsection{Two-Dimensional Gaussian}
\label{sec:2d_example}

Next, we consider cases of two-dimensional Gaussian random variables.
These examples offer much richer symmetry groups for study as well as a greater scope for variations.
We take the inertial distribution to be uniform on $\R^2$.

We start with the standard normal distribution in two dimensions,
\begin{equation}
N_{1,1}\equiv\mathcal{N}\qty(\vec{0},\mathbbm{1}_2),
\end{equation}
where $\mathbbm{1}_n$ is the $n\times n$ identity matrix.
This distribution has as its linear symmetries all rotations about the origin and all reflections about lines through the origin, which constitute the group $O(2)$.
For further exploration, we consider a two-dimensional Gaussian with covariance not proportional to the identity,
\begin{equation}
N_{1,2} \equiv \mathcal{N}\qty(\vec{0},\mqty[1&0\\0&2]).
\end{equation}
The symmetry group of this distribution is quite complicated and described below.
Among other features, it contains the Klein 4--group, $V_4 = \qty{\mathbbm{1}, -\mathbbm{1}, \sigma_3, -\sigma_3}$, for Pauli matrix $\sigma_3$.

The linear search space that preserves $\R^2$, the general affine group in two dimensions, $\operatorname{Aff}_2(\R) = \A GL_2(\R)$, has six real parameters.
Before exploring the entire space, we first examine the subspace:
\begin{align}
\label{eq:rotation}
g(X) = \mqty[c&s\\-s&c]\,X,
\end{align}
for $c, s\in \R^\times$, where $\R^\times$ is the set of non-zero real numbers.
While this is only a rotation if $c^2+s^2=1$, we want to test if a SymmetryGAN can discover this relationship starting from this more general representation.
The symmetries represented by \Eq{rotation} are a subgroup of $GL_2(\R)$: $SO(2)\times \R^+ = \ev{\theta, r|\theta\in [0, 2\pi), r\in\R^+}$, where $\R^+$ is the set of positive real numbers.
For the $ N_{1,1}$ Gaussian, this means looking for the $r = 1$ subgroup, which is indicated by the red circle in the loss landscape in \Fig{SO2_i}.
To test the SymmetryGAN, we sample the parameters $c$ and $s$ uniformly at random from $[-1, 1]^2$, and the learned $c$ and $s$ values correspond to the expected $SO(2)$ unit circle, also shown in \Fig{SO2_i}.
We repeat this exercise for the $ N_{1,2}$ Gaussian in \Fig{SO2_iii}, where the SymmetryGAN discovers the $\Z_2$ subgroup of $V_4$ generated by a rotation by $\pi$.

\begin{figure*}[p]
    \centering
   \subfloat[]{\includegraphics[width=0.31\textwidth]{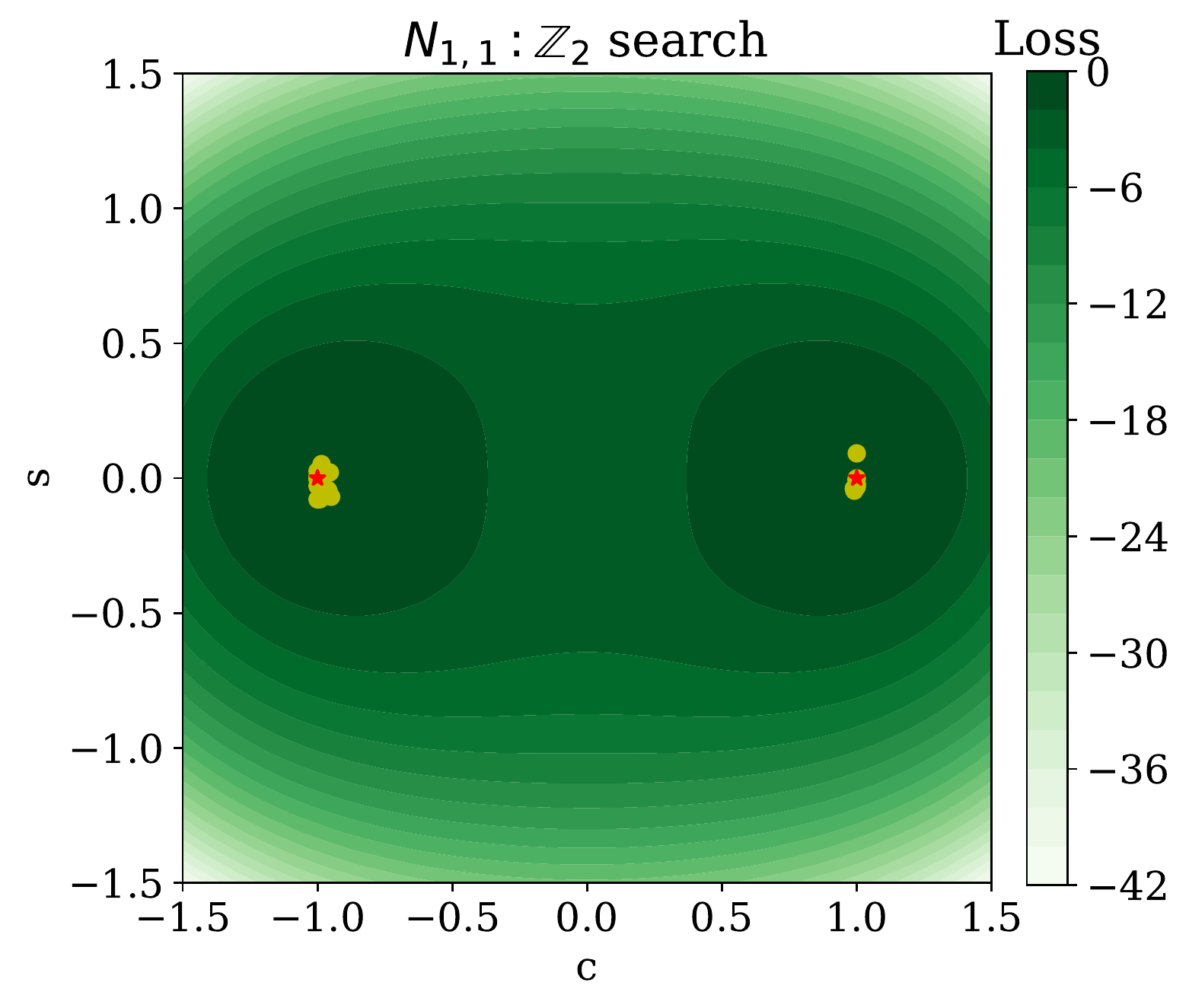}
       \label{fig:MSE_i}
       }
   $\quad$
    \subfloat[]{\includegraphics[width=0.31\textwidth]{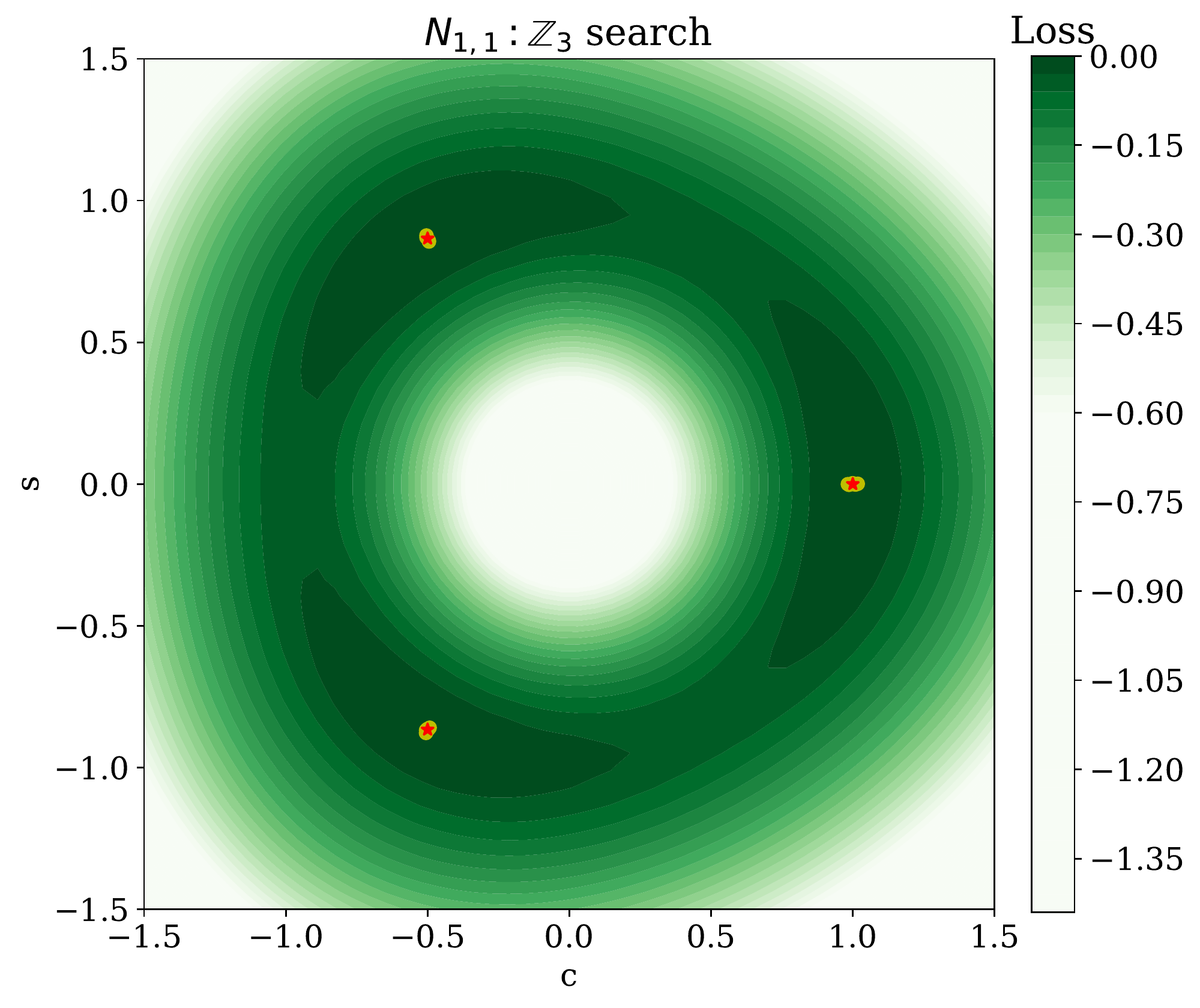}}
    $\quad$
    \subfloat[]{\includegraphics[width=0.31\textwidth]{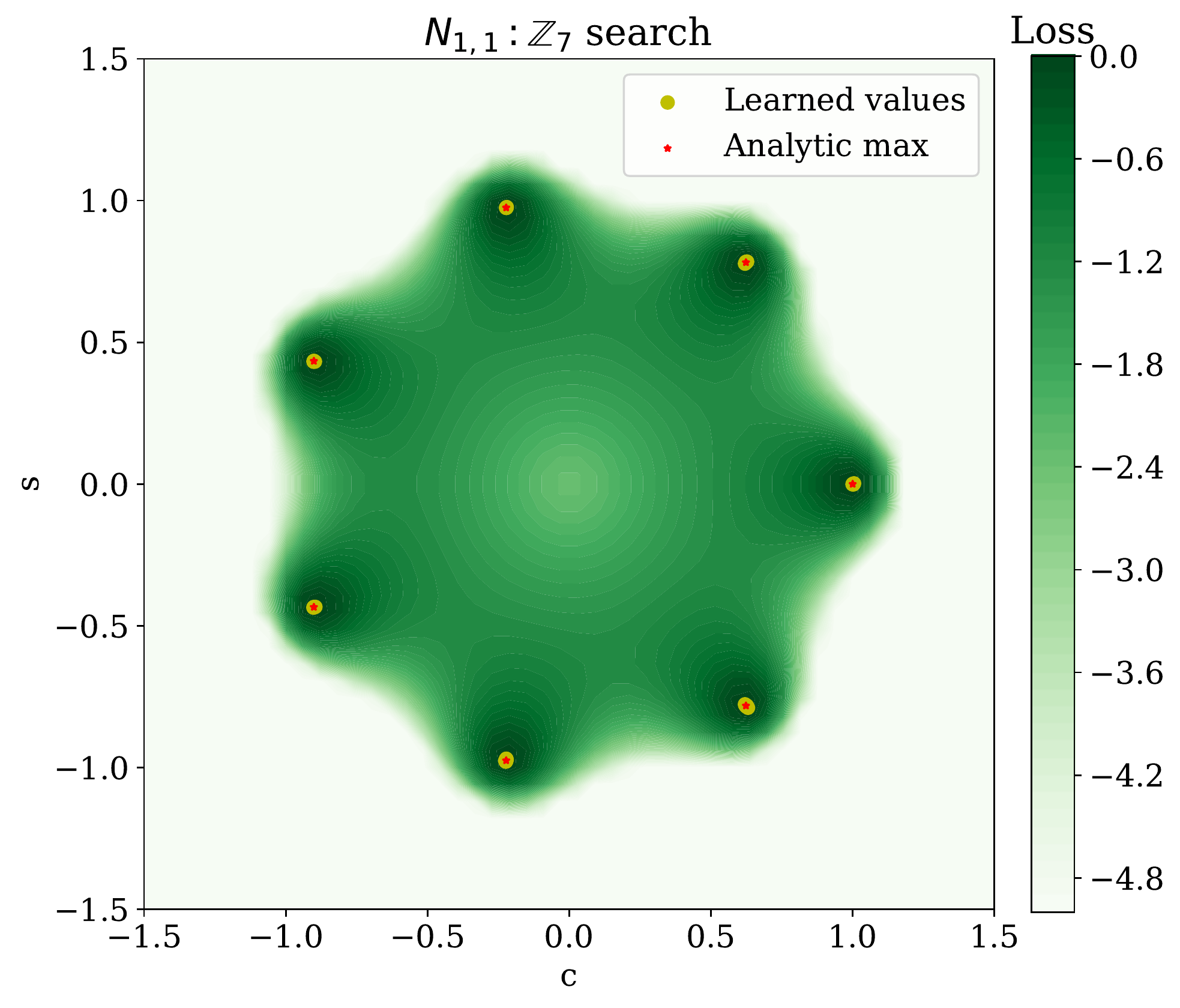}}\\

    \caption{The analytic loss landscapes overlaid with empirically discovered symmetries for the $N_{1,1}$ example with a cyclic-enforcing term added to the loss, to be compared to \Fig{SO2_i}.
    The cases studied are (i) $\Z_2$, (ii) $\Z_3$, and (iii) $\Z_7$.}
    \label{fig:MSE}
\end{figure*}

\begin{figure*}[p]
    \centering
    \subfloat[]{\includegraphics[width=0.31\textwidth]{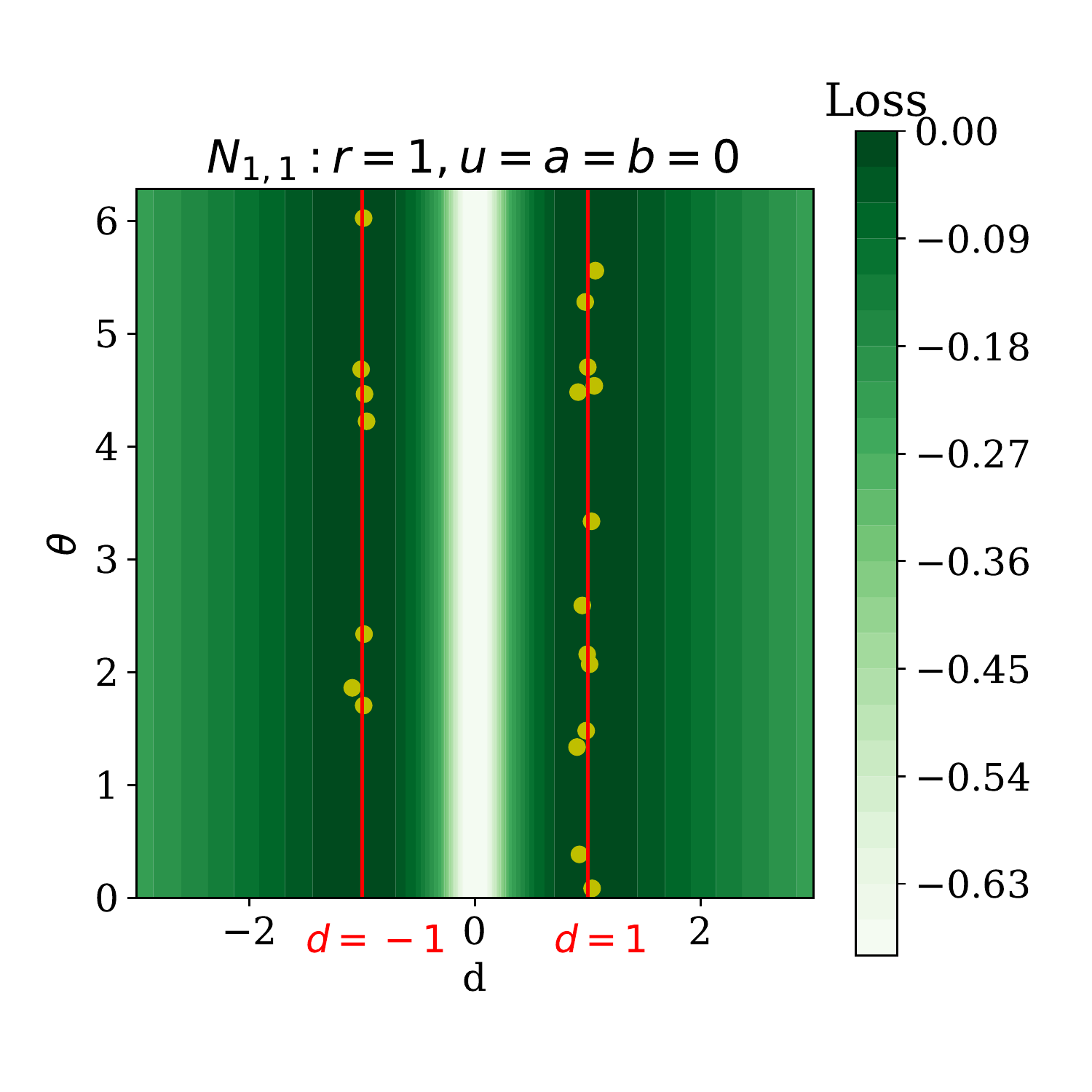}} $\quad$
     \subfloat[]{\includegraphics[width=0.31\textwidth]{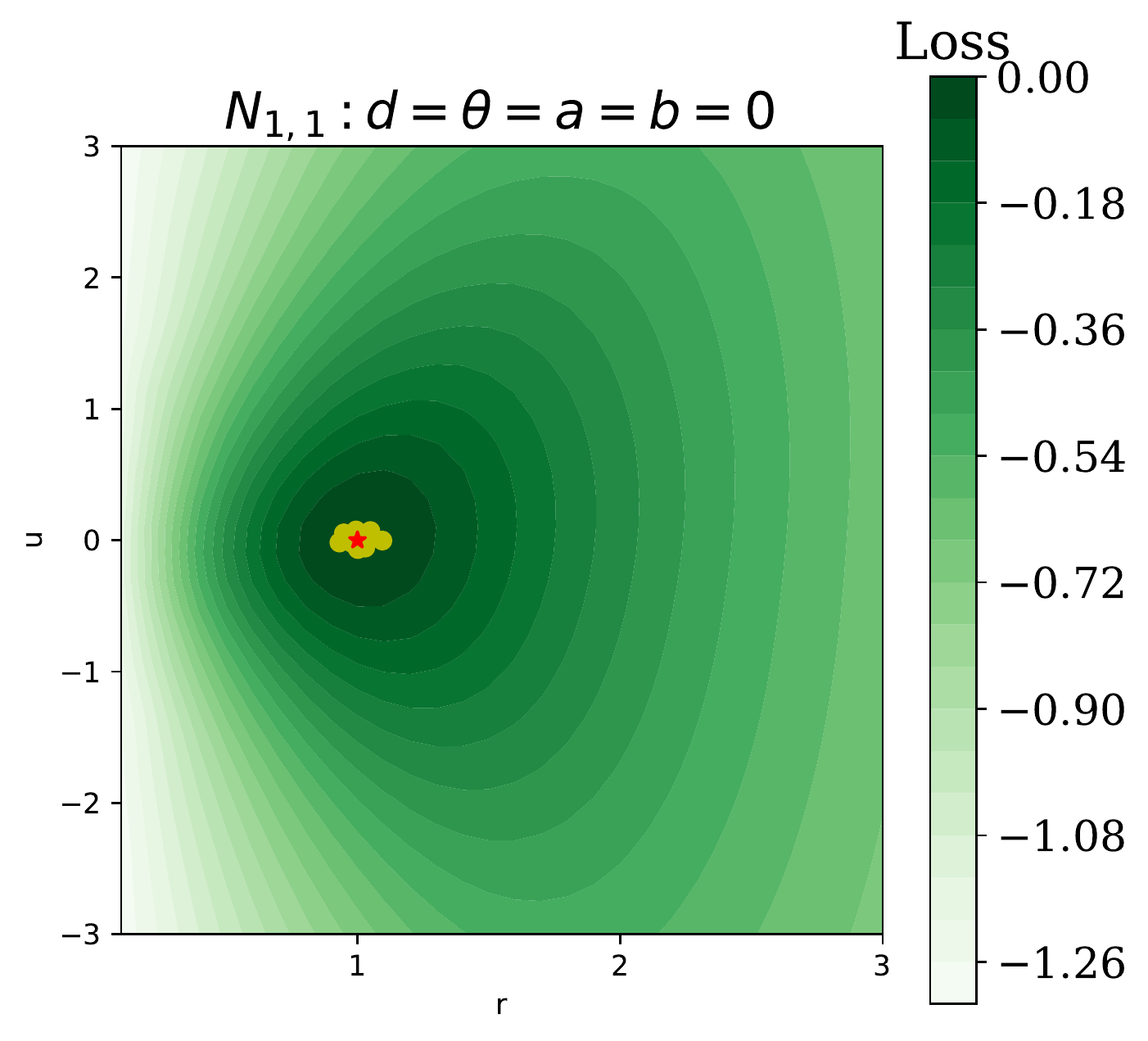}} $\quad$
  \subfloat[]{\includegraphics[width=0.31\textwidth]{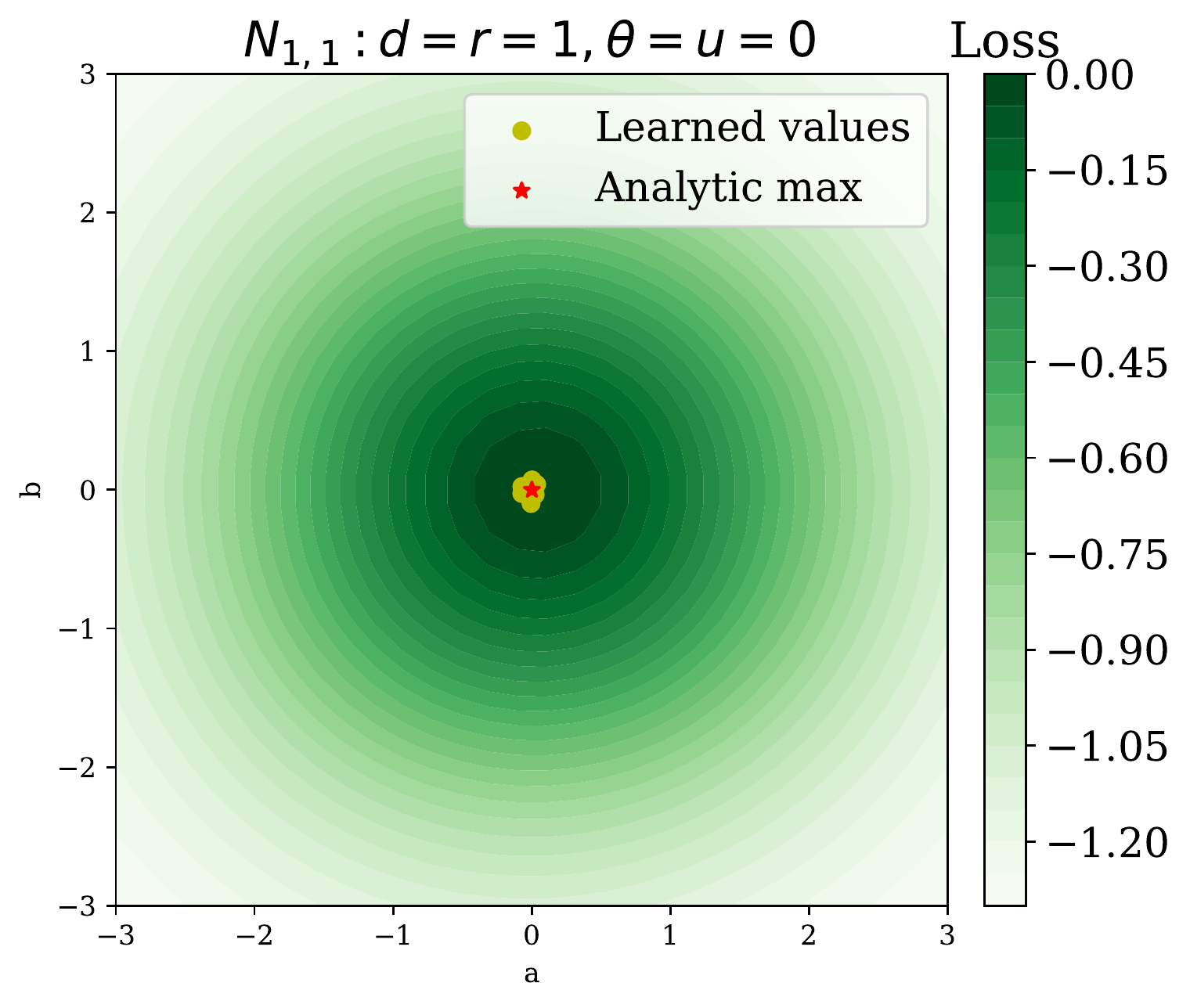}}\\
  \caption{
    Slices through the analytic loss landscape together with empirically discovered symmetries for $\mathcal N_{1,1}$ with the full $\A GL_2(\R)$ search space.
    (i) The determinant-rotation angle space.  The maxima are indicated by vertical red lines.
    (ii) The dilatation-shear space.  The maximum is indicated by a red star.
    (iii) The affine translation space.  The maximum is indicated by a red star at the origin.}
    \label{fig:AGL2symm}
\end{figure*}

\begin{figure*}[p]
    \centering
    \subfloat[]{\includegraphics[width=0.31\textwidth]{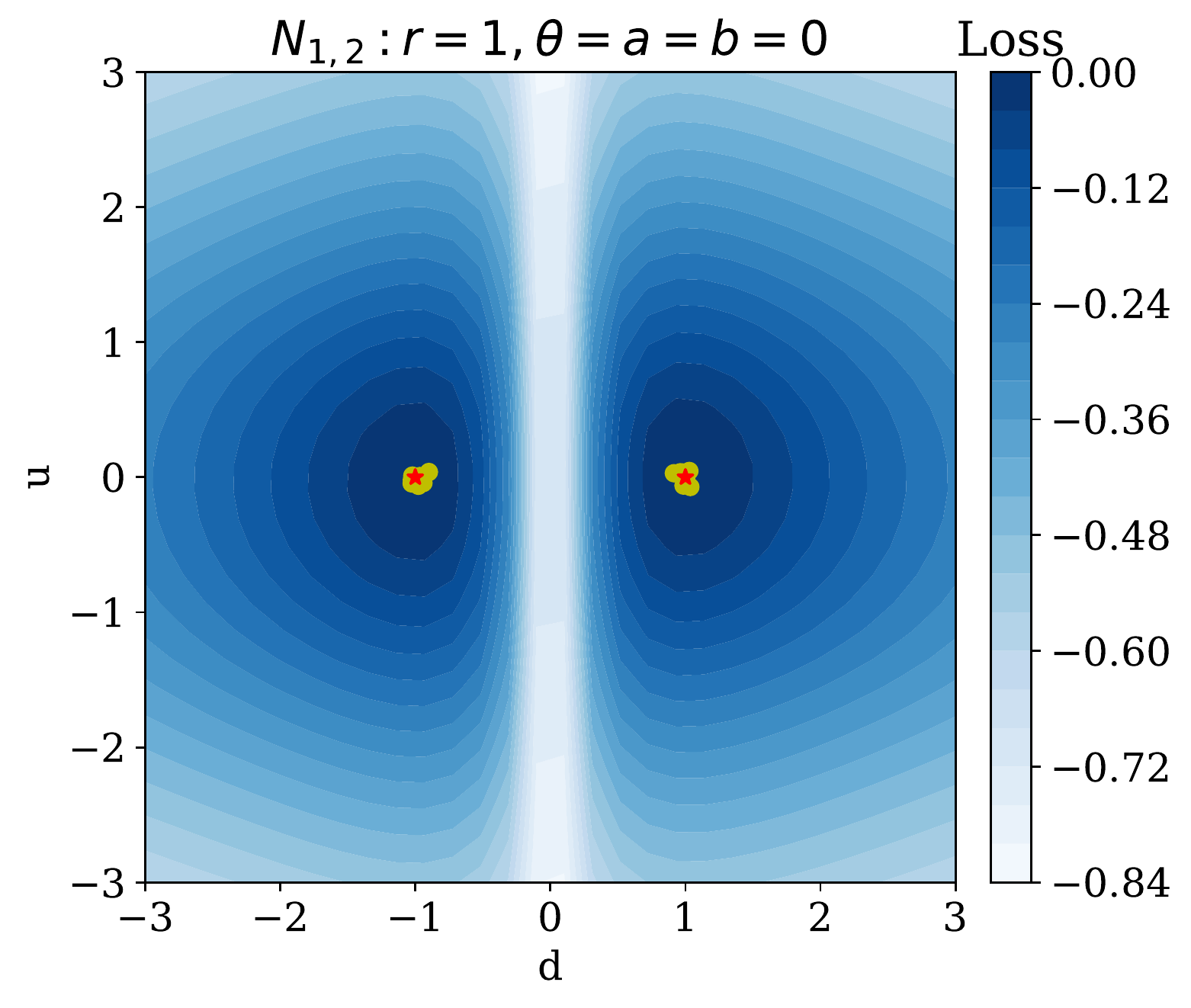}}
    $\quad$
    \subfloat[]{\includegraphics[width=0.31\textwidth]{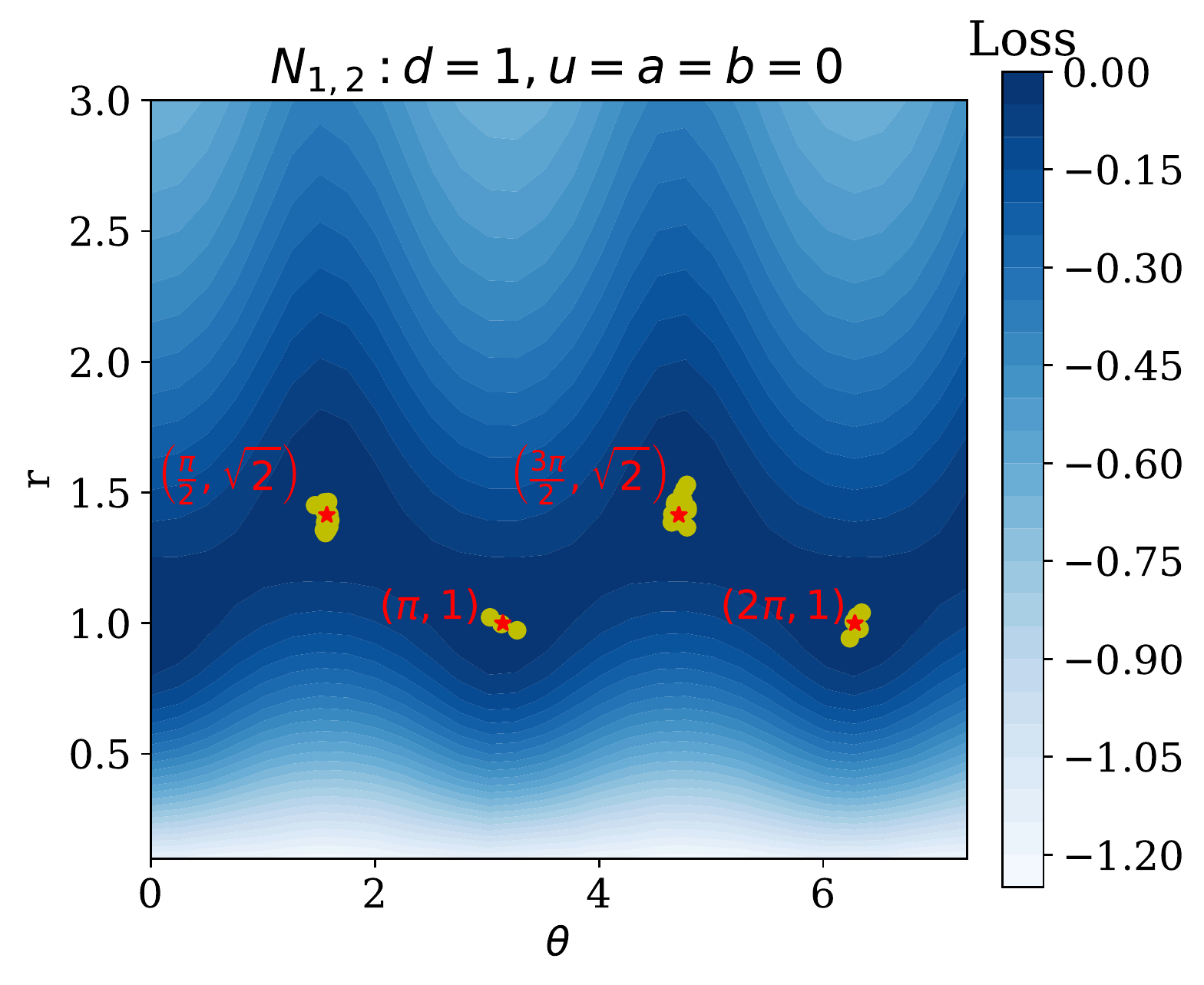}}
    $\quad$
    \subfloat[]{\includegraphics[width=0.31\textwidth]{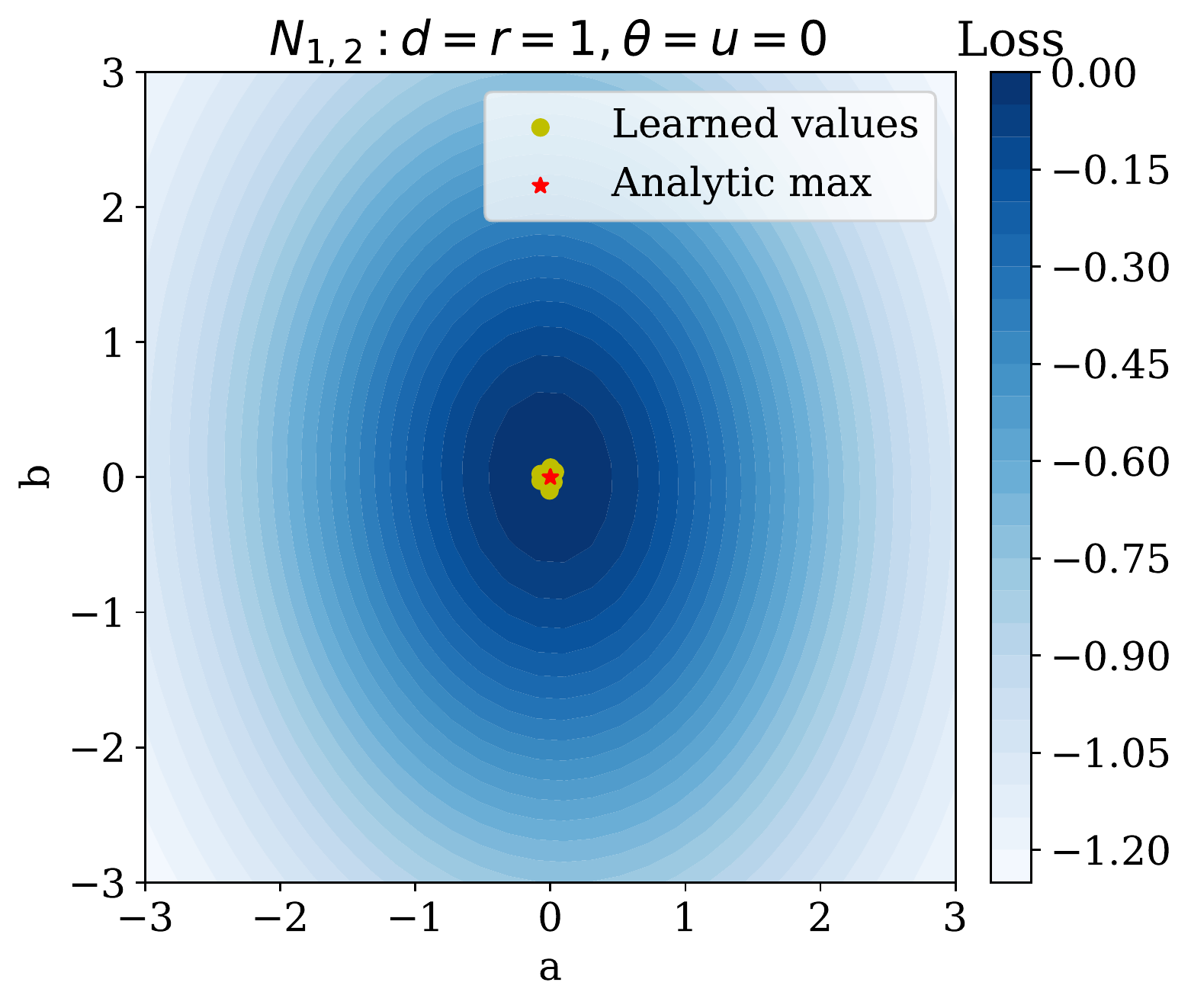}}\\
    
    \caption{
    Similar to \Fig{AGL2symm} but for the $N_{1,2}$ distribution.
    (i) The determinant-sheer space.  The maxima are indicated by two red stars.
    (ii) The dilatation-rotation angle space.  The maxima are indicated by four red stars.
    (iii) The affine translation space.  The maximum is indicated by a red star at the origin.}
    \label{fig:AGL2asymm}
\end{figure*}

This two-dimensional example allows us to test the approach in \Eq{cyclicloss} for finding $\Z_q$ subgroups of the full symmetry group.
Restricting our attention to the $N_{1, 1}$ example and the $SO(2)\times \R^+$ subgroup in \Eq{rotation}, we add the cyclic-enforcing mean squared error term to the loss with $\alpha=0.1$.
Results are shown in \Fig{MSE} for $q = 2$, $3$, and $7$, where the analytic loss optima and empirically found symmetries are broken into discretely many solutions, with the number corresponding to the $q^{\textrm{th}}$ roots of unity, as expected.

We now consider the general affine group, $\operatorname{Aff}_2(\R)$.
In two dimensions, the elements of this group can be represented as a matrix with 6 parameters:
\begin{itemize}
    \item $d\in\R^\times$, the determinant;
    \item $\theta\in [0, 2\pi)$, the angle of rotation;
    \item $r\in \R^+$, the dilatation;
    \item $u\in \R$, the shear in the $x$ direction; and
    \item $(a, b)\in \R^2$ the overall affine shift.
\end{itemize}
By Iwasawa's decomposition~\cite{10.2307/1969548}, the full transformation can be written as
\begin{align}
&g(X) =\sqrt{|d|}\,\mqty[1&0\\0&-1]^{\delta}\mqty[c_\theta&s_\theta\\-s_\theta&c_\theta]\mqty[r&0\\0&\frac{1}{r}]\mqty[1&u\\0&1]\,X+\mqty[a\\b]\,,
\end{align}
where $\delta=\frac{1 - \operatorname{sgn}(d)}{2}$ and $c_\theta=\cos(\theta)$ and $s_\theta=\sin(\theta)$.

For the distribution $N_{1,1}$, the symmetry group is $O(2)$, described by the parameters $d = \pm 1, \theta\in[0, 2\pi), r = 1$, and $u = a = b = 0$.
Visualizing this space is difficult, but multiple slices through the analytic loss landscape are presented in \Fig{AGL2symm}.
The neural network is trained over all six parameters of the Iwasawa decomposition of $\operatorname{Aff}_2(\R)$.
The empirically discovered symmetries, shown as yellow dots in \Fig{AGL2symm}, are two-parameter slices of the discovered symmetry group, where slices are chosen such that the parameters not under study are closest to $d = r = 1$, $\theta = a = b = 0$.
The empirical data agree well with the predictions.

The same analysis of $N_{1,2}$ is more complex because the corresponding symmetry group is more complicated than for $N_{1,1}$.
When $r = 1$ and $u = 0$, the symmetries are the $V_4$ we saw earlier ($\theta = 0,\pi$ and $d = \pm 1$).
By varying $r$ and $u$, however, one can in fact undo the symmetry breaking induced by the non-identity covariance, thereby restoring the rotational symmetry.
For example, when $r = \sqrt{2}$, $N_{1,2}$ is transformed into a Gaussian with covariance $\mathrm{diag}[2,1]$, therefore $r = \sqrt 2$ and $\theta = \frac\pi2, \frac{3\pi}2$ constitutes a symmetry.
It is difficult to describe the whole symmetry group in closed form, or even to visualize it because it does not live in any single planar slice of $\A GL_2(\R)$.
As shown for various parameter slices in \Fig{AGL2asymm}, though, the empirical results agree well with the analytic predictions.

\begin{figure*}[p]
    \centering
   \subfloat[]{
        \includegraphics[height=0.4\textwidth]{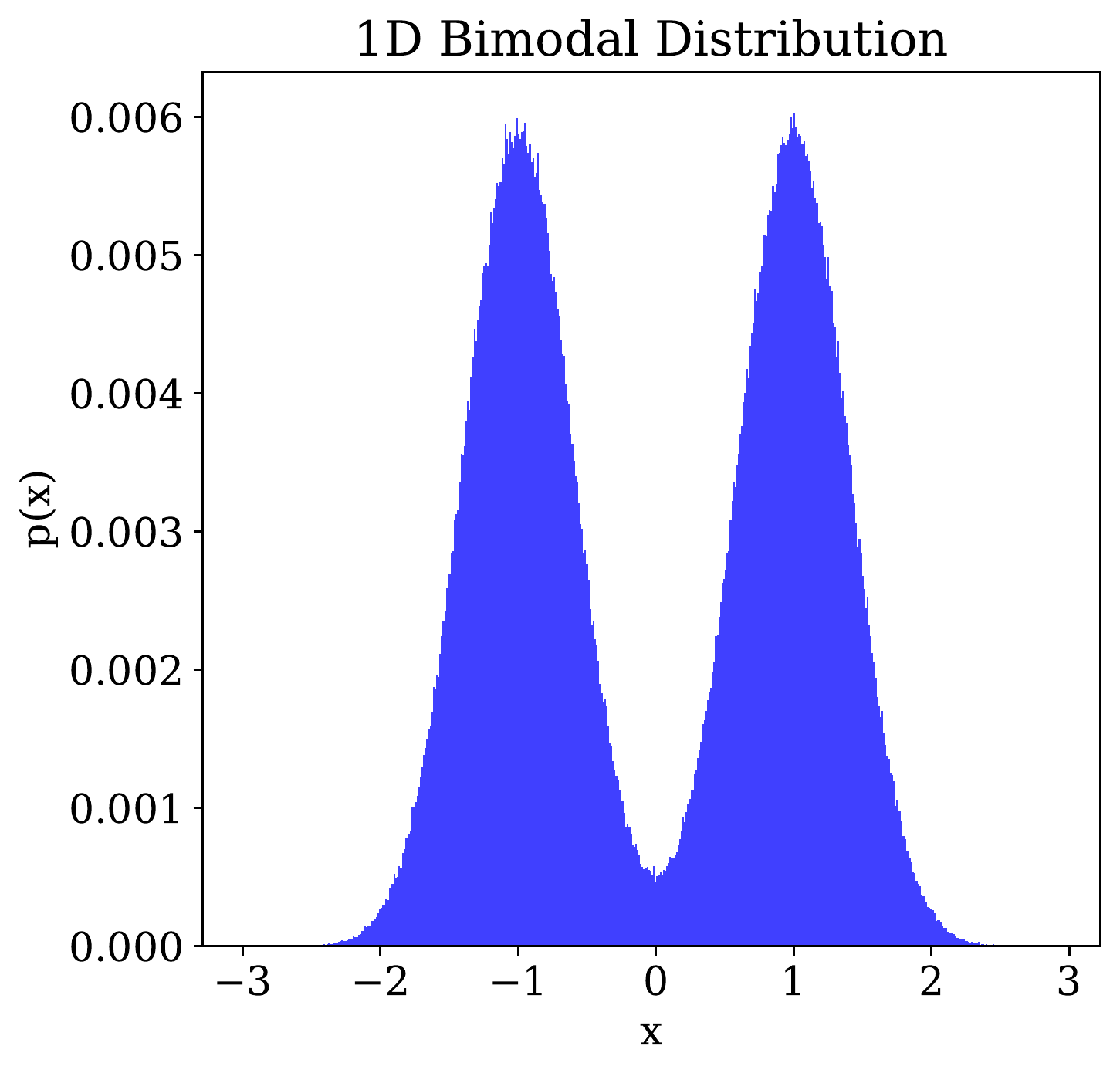}
        \label{fig:otherdistributions_1Di}
    }
    \subfloat[]{
        \includegraphics[height=0.4\textwidth]{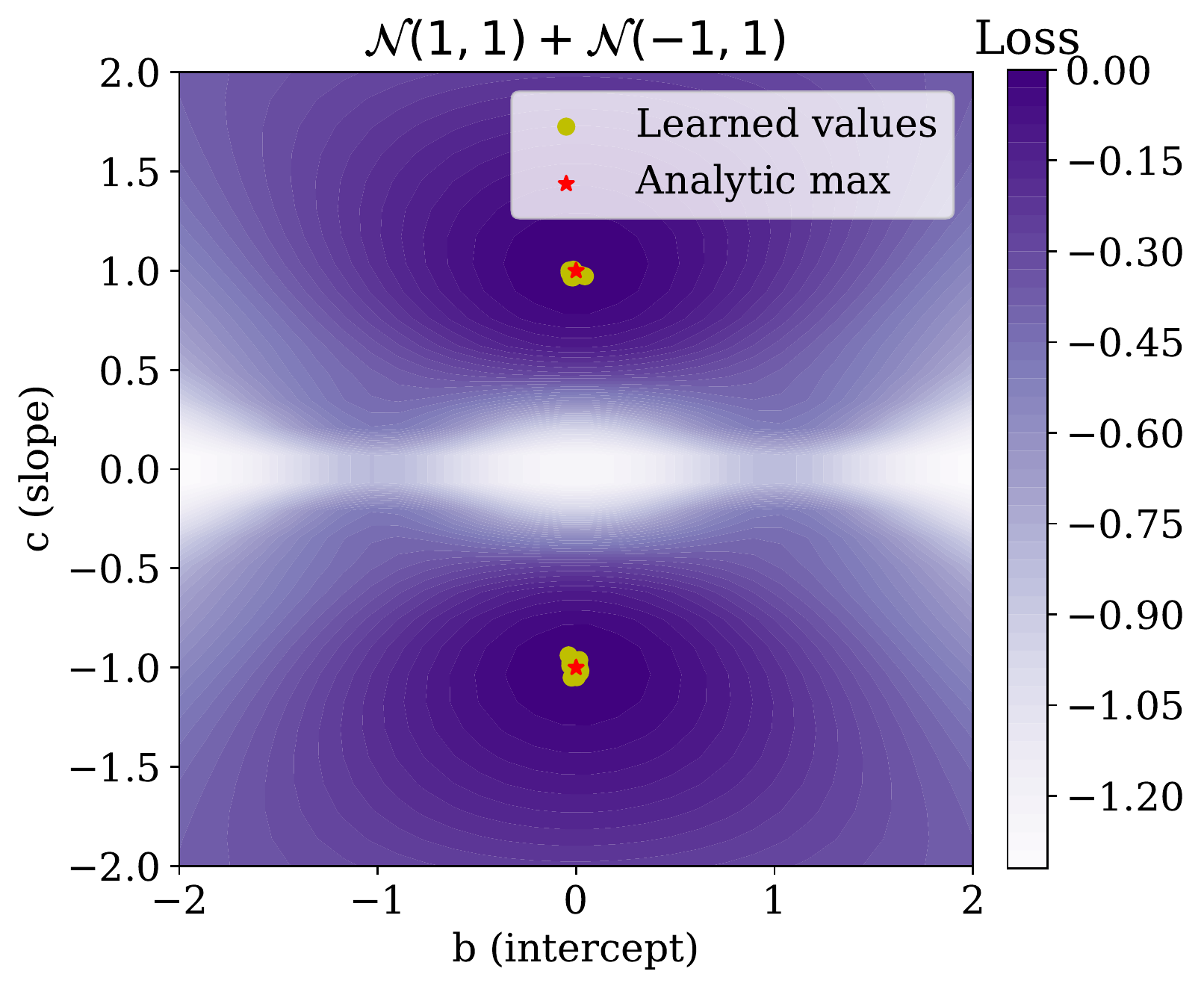}
         \label{fig:otherdistributions_1Dii}
    }
    \caption{
    Empirical distribution (i) and empirically discovered symmetries overlaid on the analytic loss landscape (ii) for a one-dimensional bimodal distribution inspired by \Ref{fisher2018boltzmann}.
    }
    \label{fig:otherdistributions_1D}
\end{figure*}

\begin{figure*}
    \subfloat[]{\includegraphics[height=0.27\textwidth]{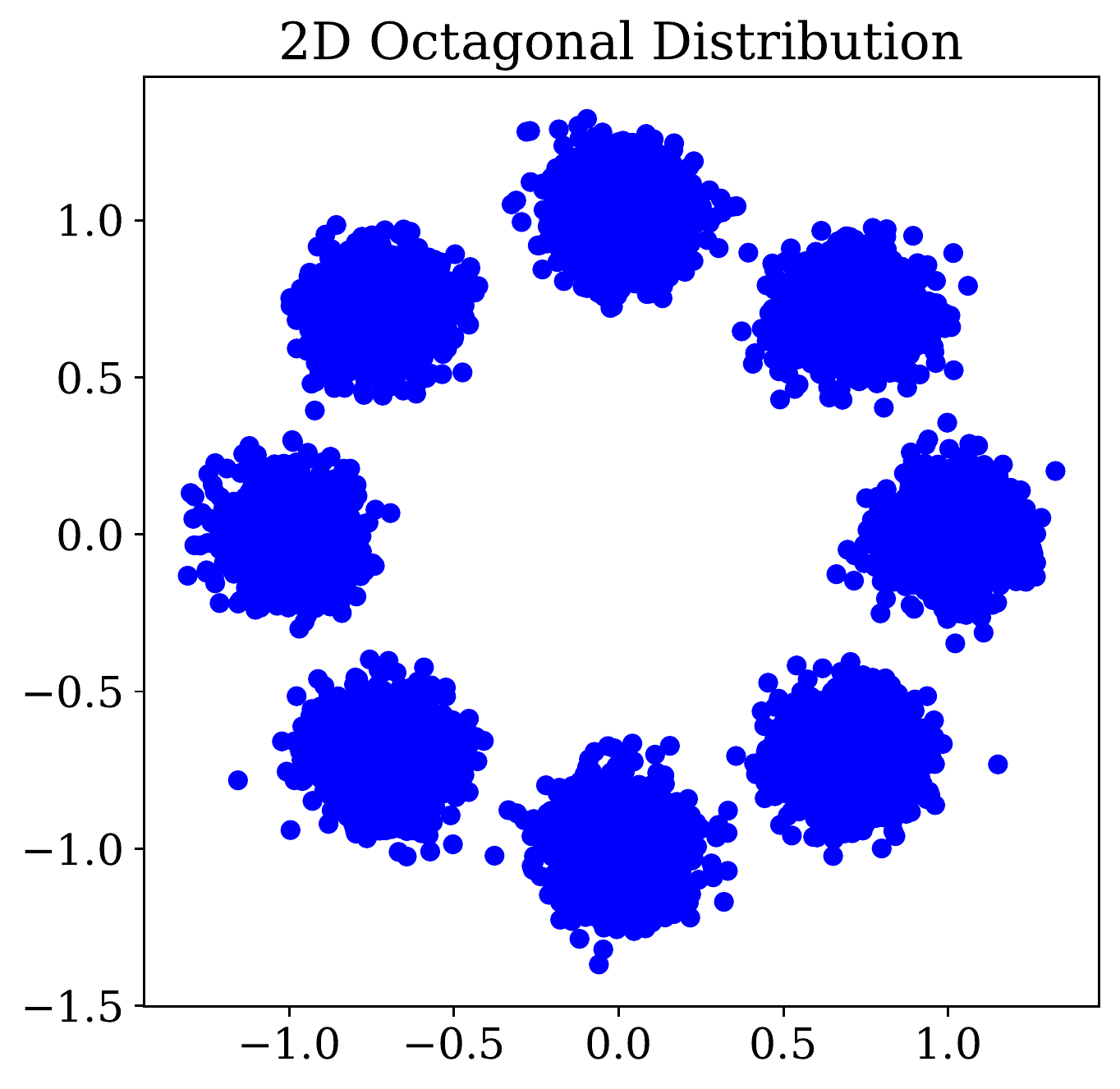}}\subfloat[]{ \includegraphics[height=0.28\textwidth]{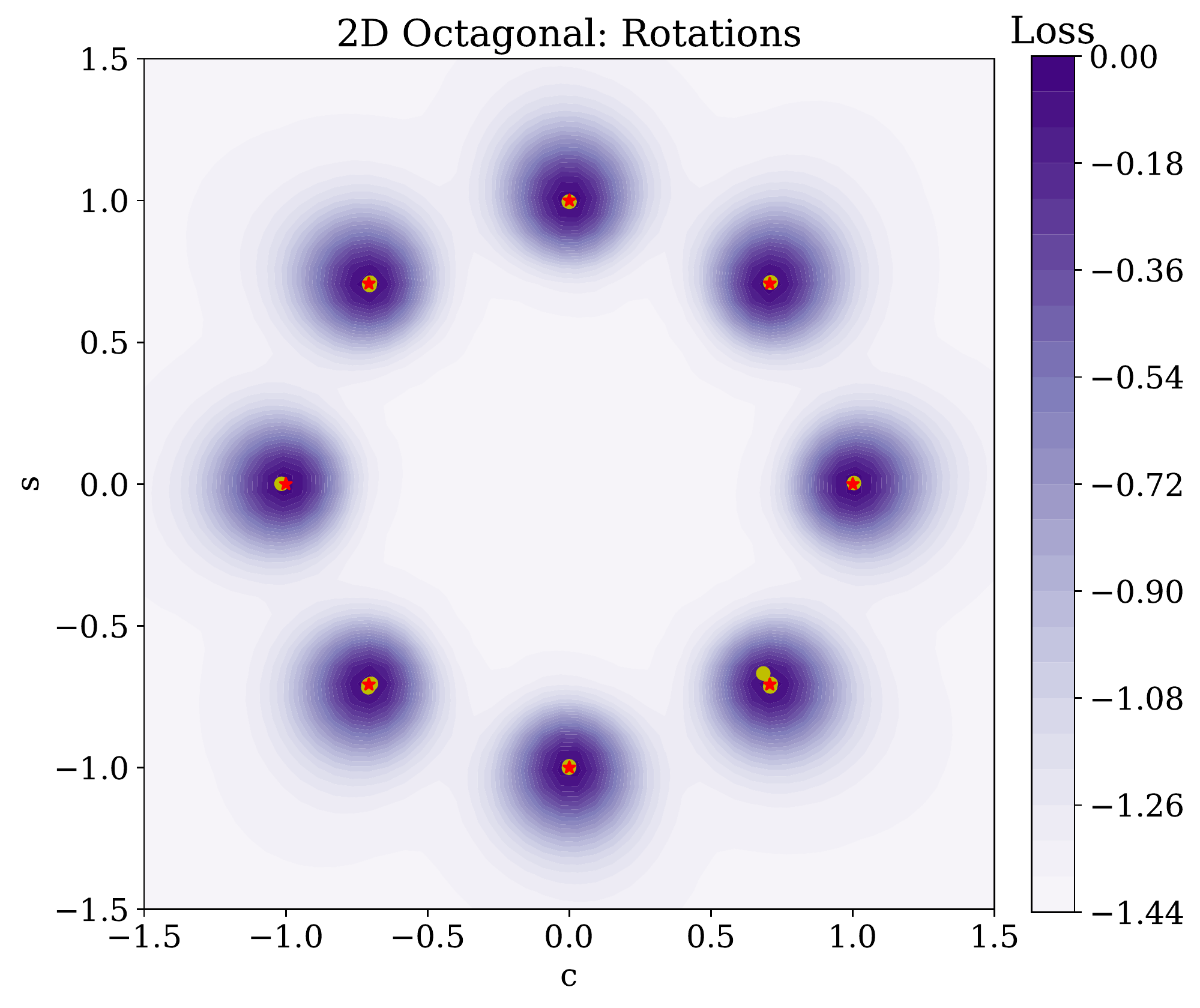}}\subfloat[]{ \includegraphics[height=0.28\textwidth]{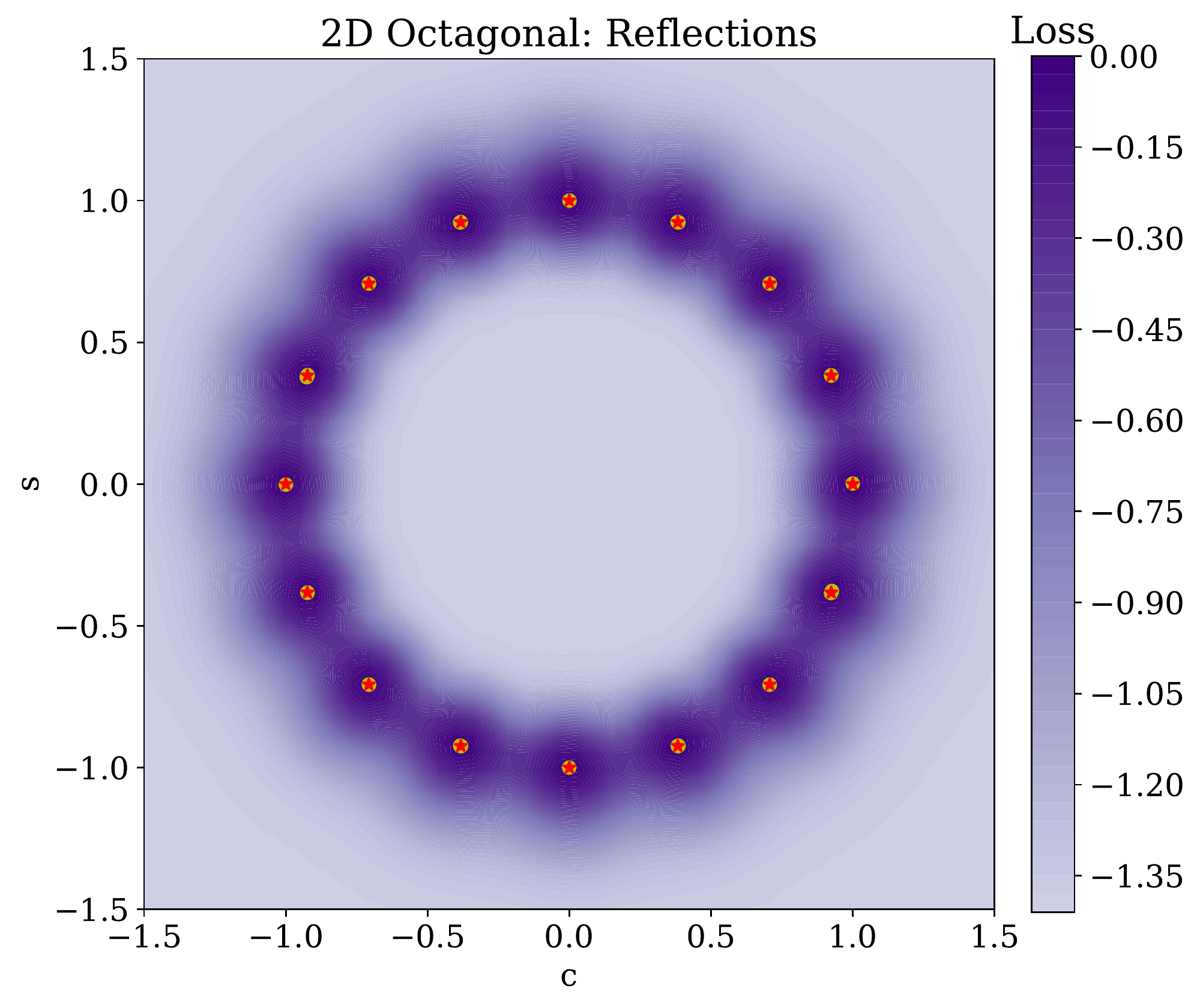}}\\
    \subfloat[]{\includegraphics[height=0.27\textwidth]{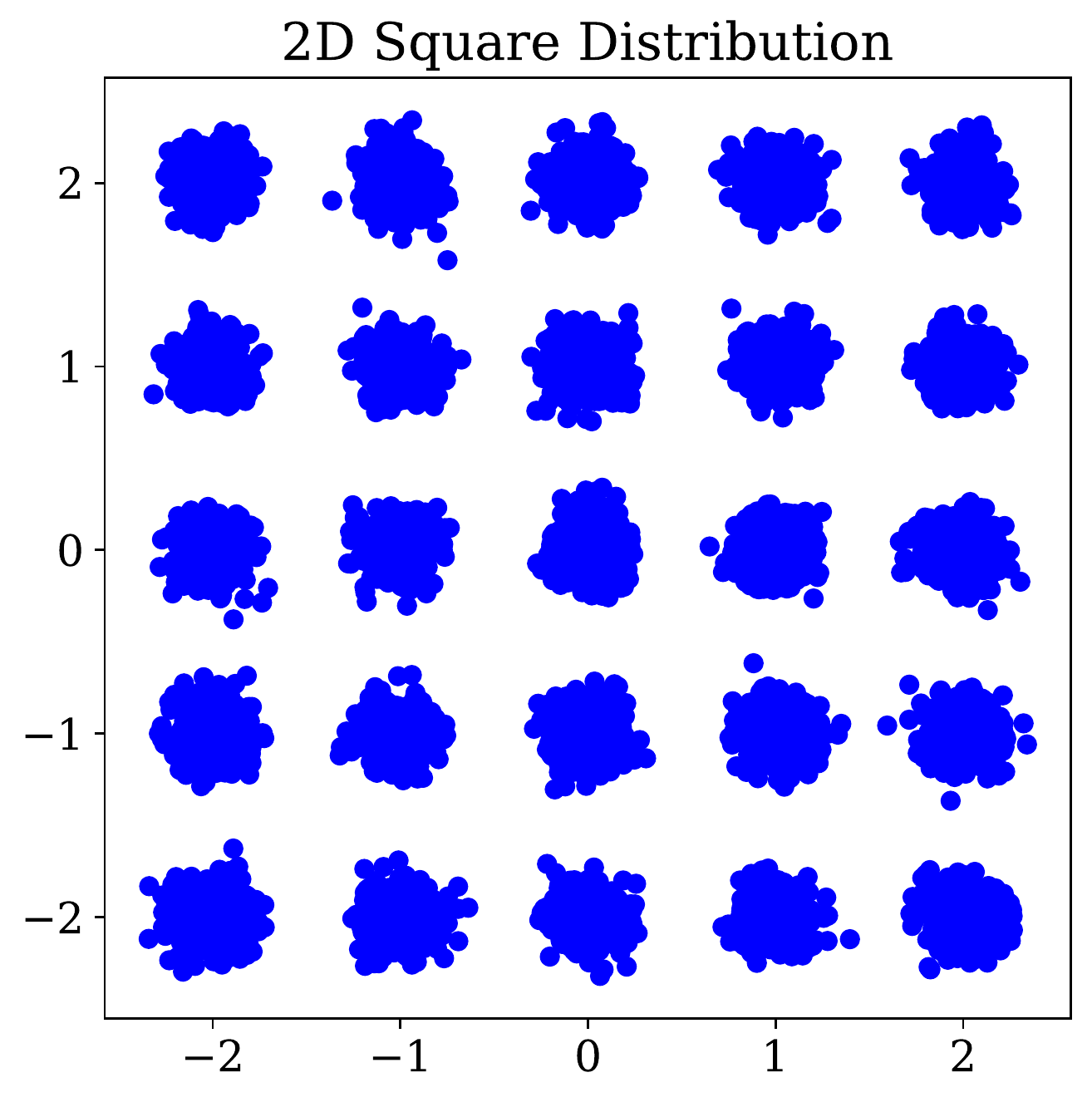}}\subfloat[]{\includegraphics[height=0.28\textwidth]{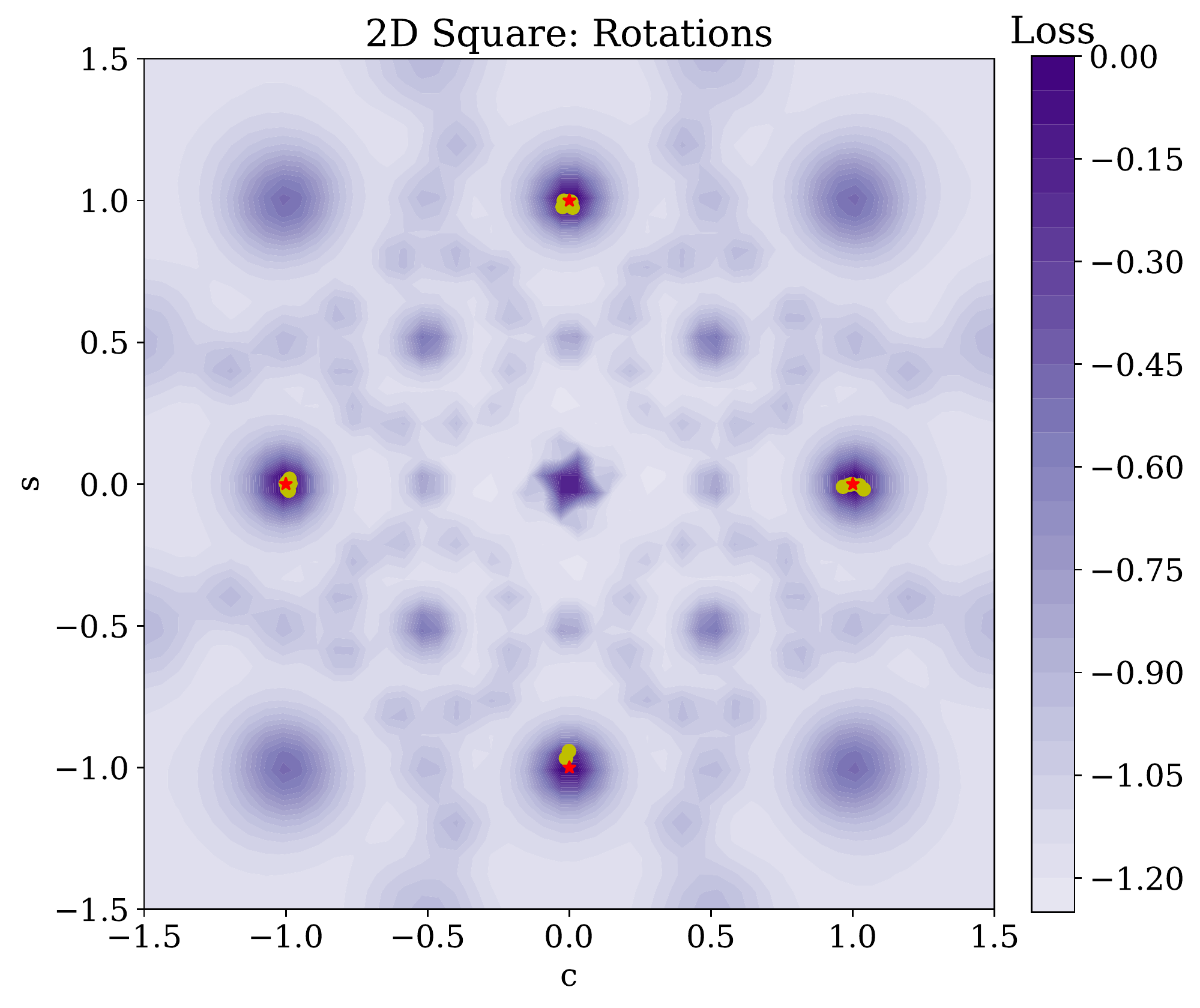}}\subfloat[]{\includegraphics[height=0.28\textwidth]{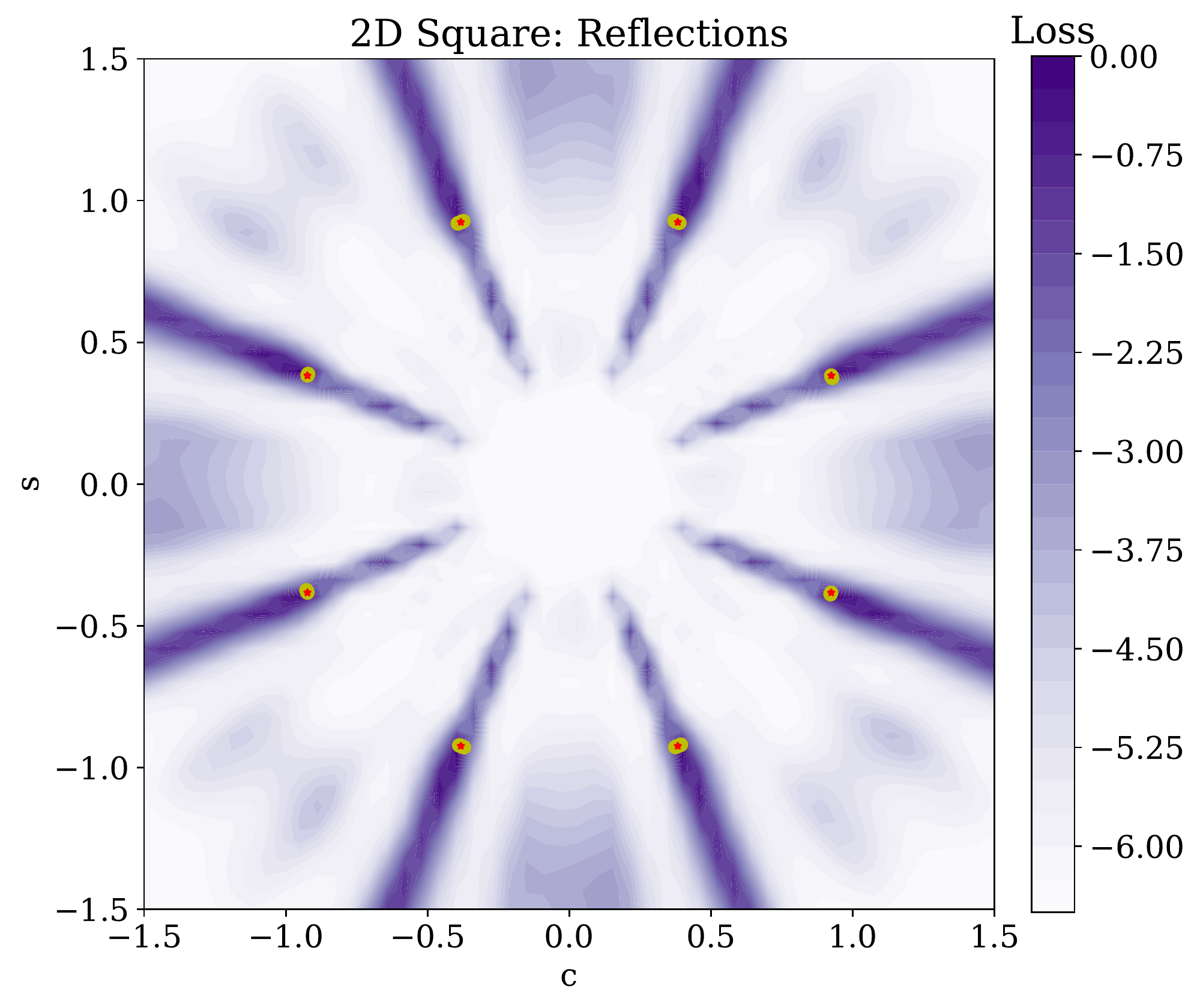}}
    \caption{
    Empirical distributions (left column) and empirically discovered rotations (middle column) and reflections (right column) overlaid on the analytic loss landscape for two two-dimensional Gaussian mixture models inspired by \Ref{fisher2018boltzmann}.
    The studied examples are (i,ii,iii) a two-dimensional octagonal distribution, and (iv,v,vi) a two-dimensional 5$\times$5 distribution. Note that antipodal points on (iii) and (vi) represent the same reflection}
    \label{fig:otherdistributions_2D}
\end{figure*}

\subsection{Gaussian Mixtures}

As our last set of simple examples, we apply the SymmetryGAN approach to three Gaussian mixture models, inspired by the examples in \Ref{fisher2018boltzmann}.
The first is a one-dimensional bimodal probability distribution:
\begin{align}
p(x) = \frac12\mathcal N (-1, 1) + \frac12 \mathcal N (1, 1),
\end{align}
which respects the $\Z_2$ symmetry group $g(x) = \pm x$.
The empirical distribution for this example is shown in \Fig{otherdistributions_1Di}.
Applying SymmetryGAN starting from the generator for linear transformations in \Eq{linear_form}, it finds the predicted symmetries with great accuracy, as shown in \Fig{otherdistributions_1Dii}.

We next consider two two-dimensional Gaussian mixtures.
The octagonal distribution,
\begin{align}
p(x) = \frac18\sum_{i = 1}^8\mathcal N\left(\cos\frac{2\pi i}{8}, 0.1\right)\times \mathcal N\left(\sin\frac{2\pi i}{8}, 0.1\right),
\end{align}
has the dihedral symmetry group of an octagon $D_{8}$.
The two-dimensional $5\times 5$ square distribution,
\begin{align}
p(x) = \frac1{25} \sum_{i = 1}^5\sum_{j = 1}^5\mathcal N (i - 2 , 0.1)\times\mathcal N (j-2, 0.1),
\end{align}
has the symmetry group of a square $D_4$.
We use the generator \begin{equation}
    \label{eq:O2}
    g(X) = \mqty[c&s\\-s&(-1)^\delta c]X,
\end{equation}
which can discover the the entire symmetry subgroup (rotations and reflections) in $O(2)$.
Data sampled from these distributions are shown in the left column of \Fig{otherdistributions_2D}.
In the middle and right columns of \Fig{otherdistributions_2D}, we see that SymmetryGAN finds the expected rotations and reflections, respectively.

\section{Particle Physics Example}
\label{sec:hepexample}

We now turn to an application of SymmetryGANs in particle physics.
Here, we are interested to learn if this approach can recover well-known azimuthal symmetries in collider physics and possibly identify symmetries that are not immediately obvious.
By the Coleman--Mandula theorem~\cite{PhysRev.159.1251}, space-time and internal symmetries cannot be combined in any but a trivial way.
Ergo, from momentum data, the only symmetry groups that can be discovered are subgroups of the Poincar\'e group, $\R^{1, 3}\rtimes O(1, 3)$.
There is much to be explored and studied within the Poincar\'e group itself, however.
We do not even have a complete classification of its unitary representations~\cite{doi:10.1142/0537, 10.2307/1968551} and its subgroup structure is remarkably rich and complex.
Discovering which specific subgroup of the Poincar\'e group constitutes the symmetry group of the system at hand is a non-trivial question, one we can seek to address through SymmetryGAN.

\subsection{Dataset and Preprocessing}

This case study is based on dijet events.
Jets are collimated sprays of particles produced from the fragmentation of quarks and gluons, and pairs of jets are one of the most common configurations encountered at the LHC.
With a suitable jet clustering algorithm, each jet has a well-defined momentum, and we can search for symmetries of the jet momentum distributions.

The dataset we use is the background dijet sample from the LHC Olympics anomaly detection challenge~\cite{gregor_kasieczka_2019_4536377,Kasieczka:2021xcg}.
These events are generated using \texttt{Pythia} 8.219~\cite{Sjostrand:2006za,Sjostrand:2007gs} with detector simulation provided by \texttt{Delphes} 3.4.1~\cite{deFavereau:2013fsa,Mertens:2015kba,Selvaggi:2014mya}
The reconstructed particle-like objects in each event are clustered into
$R=1$ anti-$k_T$~\cite{Cacciari:2008gp} jets using \texttt{FastJet} 3.3.0~\cite{Cacciari:2011ma,Cacciari:2005hq}.
All events are required to satisfy a single $p_T>1.2$~TeV jet trigger, and our analysis is based on the leading two jets in each event, where leading refers to the ones with largest transverse momenta ($p_T^2 = {p_x^2 + p_y^2}$).

Each event is represented as a four-dimensional vector:
\begin{equation}
X = (p_{1x},p_{1y},p_{2x},p_{2y}),
\end{equation}
where $p_1$ refers to the momentum of the leading jet, $p_2$ represents the momentum of the subleading jet, and $x$ and $y$ are the Cartesian coordinates in the transverse plane.
We focus on the transverse plane because the jets are typically back-to-back in this plane as a result of momentum conservation.
The longitudinal momentum of the parton-parton interaction is not known and so there is no corresponding conservation law for $p_z$.%
\footnote{In principle, we could use SymmetryGAN to confirm the absence of a symmetry in $p_z$.}

Since we have a four-dimensional input space, a natural search space for symmetries is $SO(4)$, the group of all rotations on $\R^4$.  Before exploring the whole candidate symmetry space, we first consider an $SO(2)\times SO(2)$ subspace where the two leading jets are independently rotated.

\subsection{$SO(2) \times SO(2)$ Subspace}

\begin{figure*}[p]
    \centering
    \subfloat[]{\includegraphics[height=0.45\textwidth]{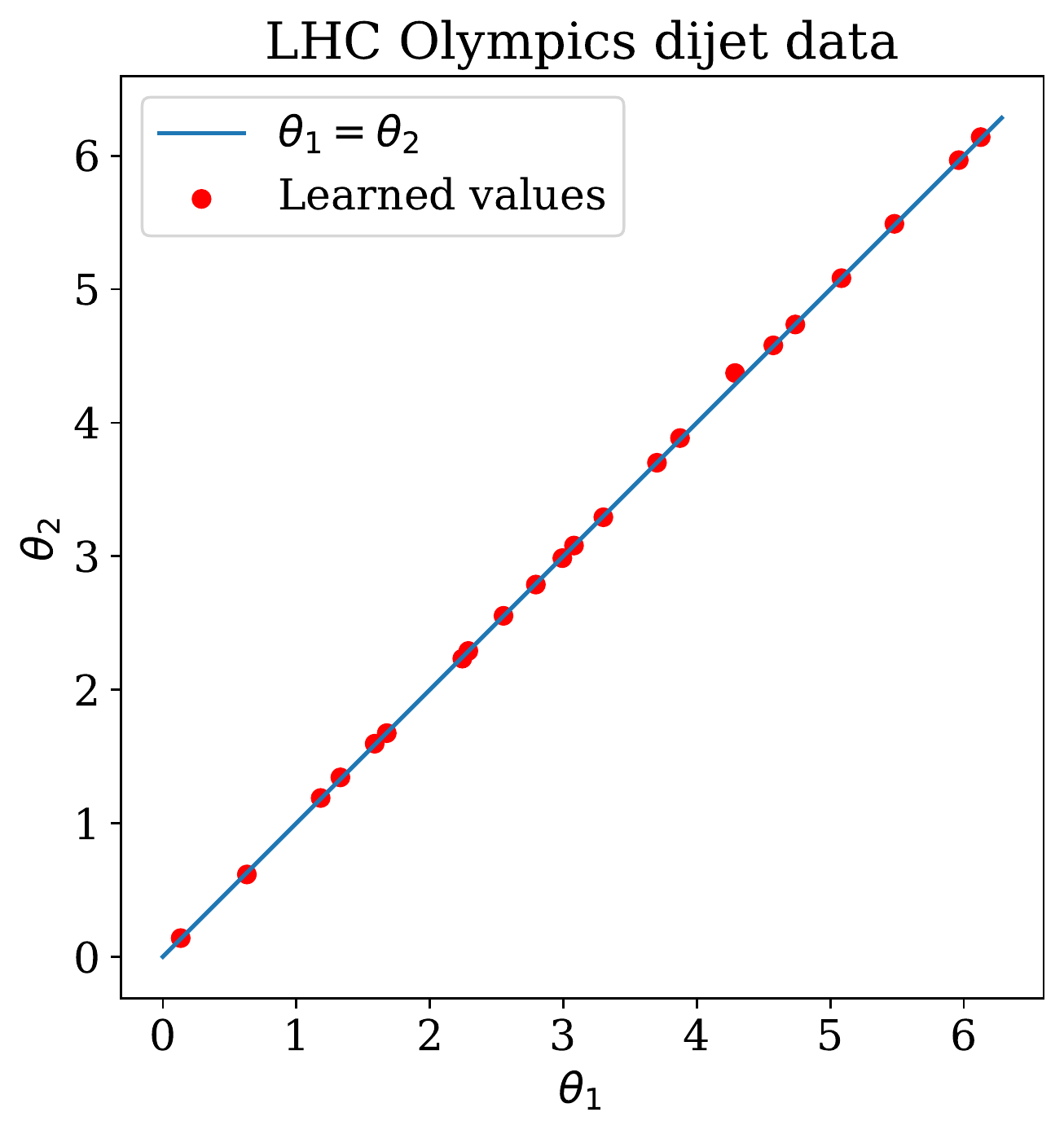}
    \label{fig:LHCO_i}}
    $\quad$
    \subfloat[]{\includegraphics[height=0.45\textwidth]{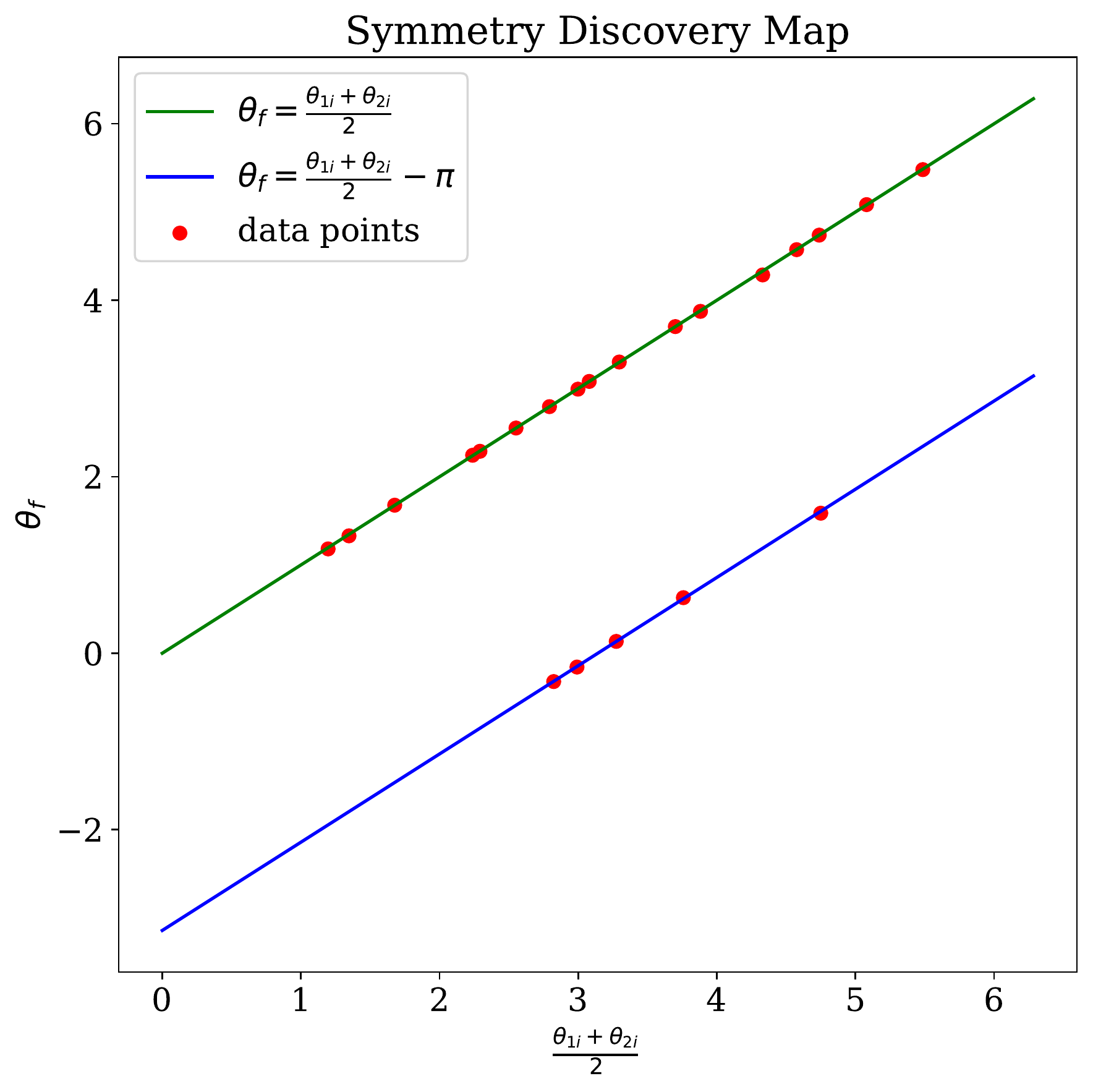}
    \label{fig:LHCO_ii}}
    \caption{
    (i) Empirically discovered symmetries in the LHC Olympics dijet dataset.
    The final values of $\theta_1$ and $\theta_2$ from the SymmetryGAN are plotted over the line $\theta_1 = \theta_2$.
    (ii) The map between initial and final symmetry parameters.
    The final rotation angle is the average of the initialized rotation angles, offset by $\pi$ if the angle between the initialized angles is reflex. 
    }
    \label{fig:LHCO}
\end{figure*}  

Because of momentum conservation, we expect that only those rotations that simultaneously rotate both jets by the same angle will be symmetries.
We start from a generic $SO(2)\times SO(2)$ group element:
\begin{equation}
    g_{\theta_1, \theta_2}\mqty[p_{1x}\\ p_{1y}\\p_{2x}\\p_{2y}] = \mqty[\cos\theta_1&\sin\theta_1&0&0\\-\sin\theta_1&\cos\theta_1&0&0\\0&0&\cos\theta_2&\sin\theta_2\\0&0&-\sin\theta_2&\cos\theta_2]\mqty[p_{1x}\\ p_{1y}\\p_{2x}\\p_{2y}],
\end{equation}
where $(\theta_1, \theta_2) \in [0, 2\pi)^2$.
We expect the symmetries to correspond to the subgroup $\qty{g_{\theta_1, \theta_2}|\theta_1 = \theta_2}\cong SO(2)$.
This prediction is borne out in \Fig{LHCO_i}.

We can also study the training dynamics of the SymmetryGAN.
More information about this procedure is given in \App{symmetry_discovery_map}, but the idea is to find a symmetry discovery map $\Omega: SO(2)\times SO(2) \to SO(2)$, $(\theta_{1i}, \theta_{2i})\mapsto\theta_{f},$ that describes how the initial parameters map to the learned ones.
We propose the map given by
\begin{equation}
\begin{split}
    \Omega(\theta_1, \theta_2) &= \begin{cases}\frac{\theta_1 + \theta_2}{2}& |\theta_1 - \theta_2| < \pi\,,\\
    \frac{\theta_1 + \theta_2}{2} - \pi & |\theta_1 - \theta_2| > \pi\,,
    \end{cases}
    \end{split}
\end{equation}
where there is only one output angle even though the output space is two-dimensional.
This map posits that the final angle will bisect the smaller angle between $\theta_1$ and $\theta_2$, which is validated by the empirical results shown in \Fig{LHCO_ii}.

\subsection{$SO(4)$ Search Space}

We now turn to the four-dimensional rotation group.
$SO(4)$ is a six parameter group, specified by $\qty{\theta_i}_{i=1}^6,$ which parametrize the six independent rotations:
\begin{align}
R_1\colon p_{1x}&\leadsto p_{1y},&
R_2\colon p_{1x}&\leadsto p_{2x},\\
R_3\colon p_{1x}&\leadsto p_{2y},&
R_4\colon p_{1y}&\leadsto p_{2x},\\
R_5\colon p_{1y}&\leadsto p_{2y},&
R_6\colon p_{2x}&\leadsto p_{2y},
\end{align}
where the notation $R: a \leadsto b$ means
\begin{align}
    R(a) &= a\cos\theta + b\sin\theta,\\
    R(b) &= b\cos\theta - a\sin\theta.
\end{align}
One way to describe a generic generator $g_{\vb*\theta}$ is
by
\begin{equation}
    g_{\vb*\theta}(X) = R_1 R_2 R_3 R_4 R_5 R_6 \, X.
\end{equation}

\begin{figure*}[p]
    \centering
        \subfloat[]{\includegraphics[width=0.45\textwidth]{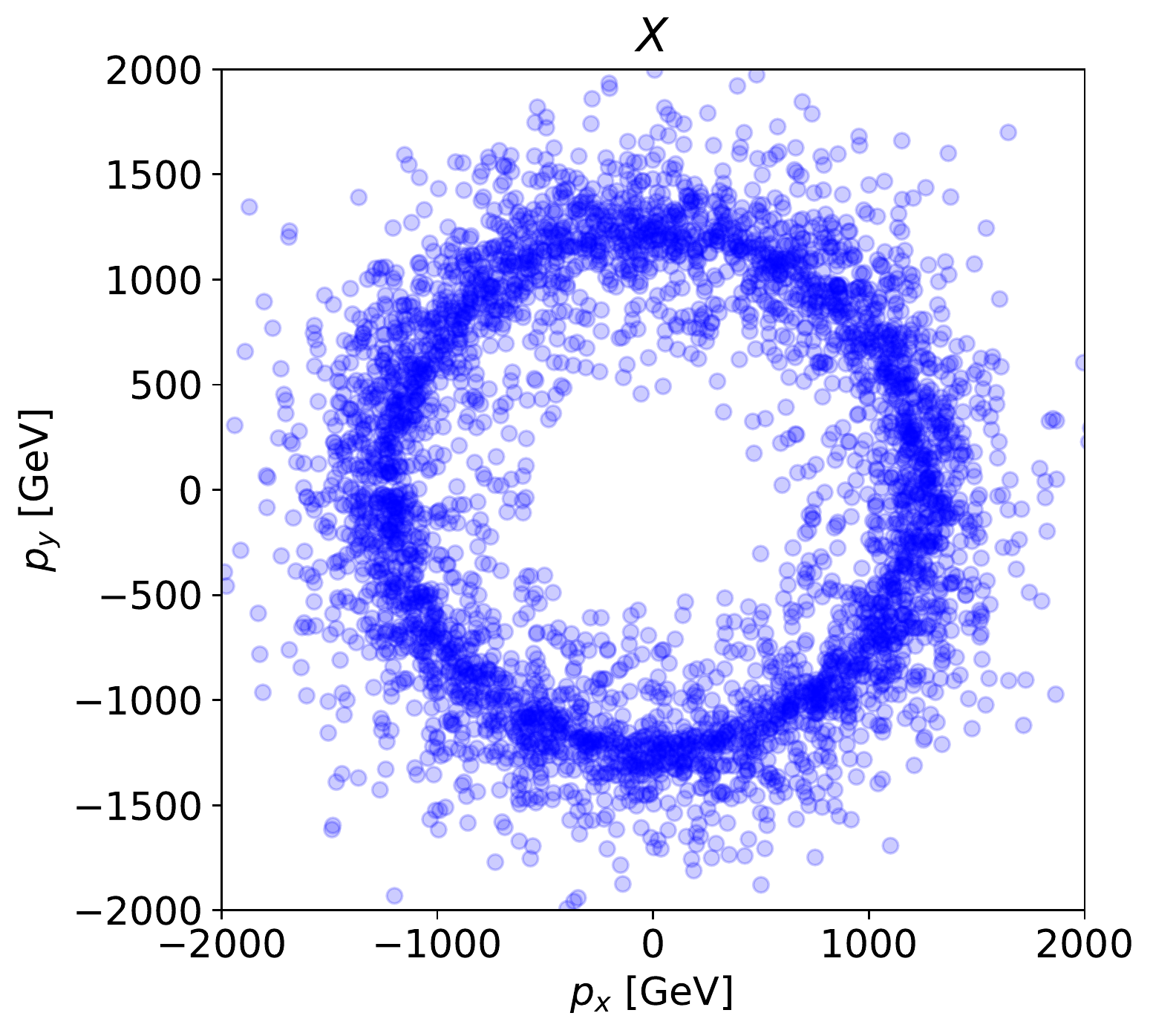} \label{fig:LHCOComparison_i}}
    $\quad$
    \subfloat[]{\includegraphics[width=0.45\textwidth]{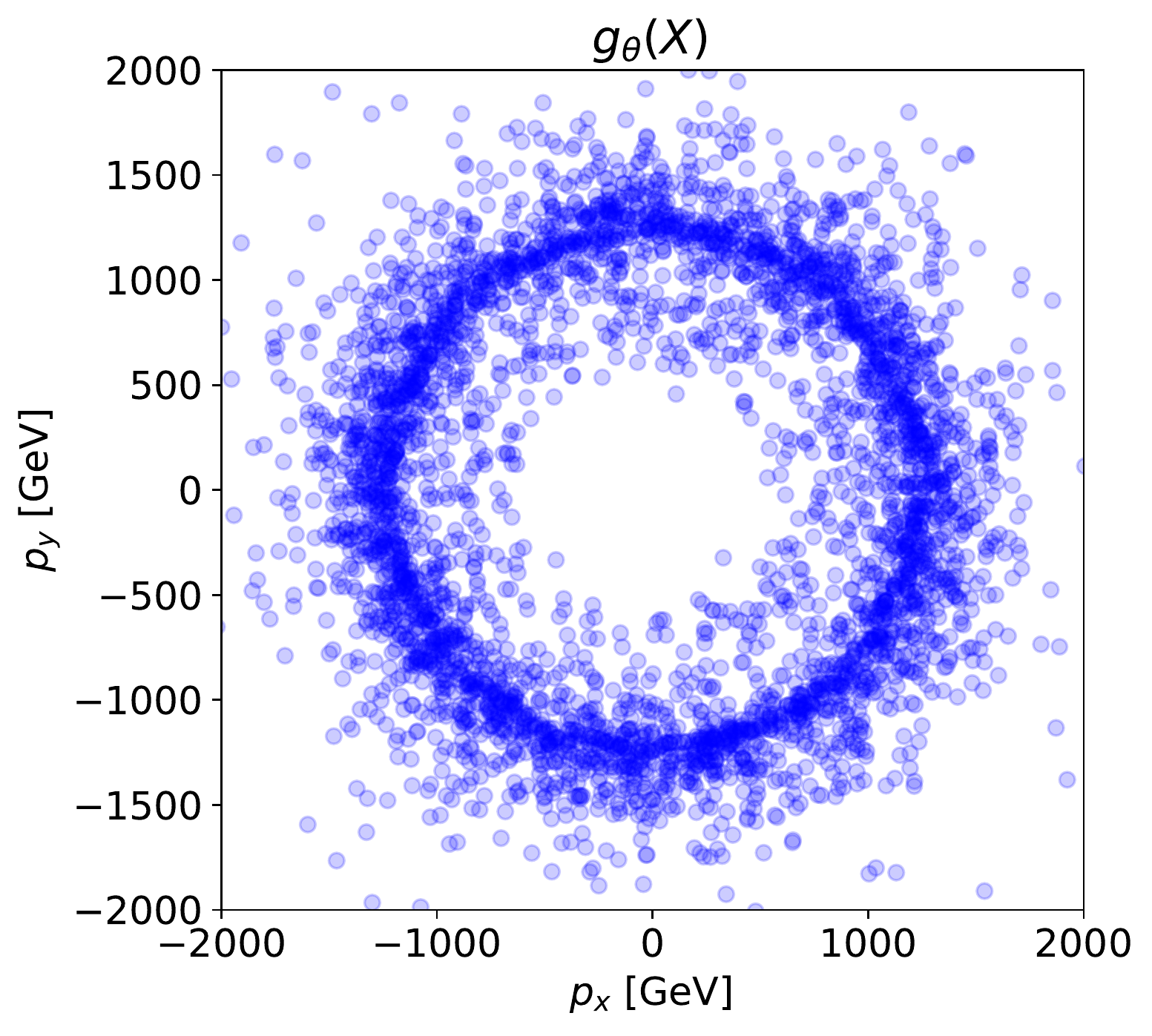}
    \label{fig:LHCOComparison_ii}}
    \caption{
    Two dimensional projection of (i) the original LHC Olympics dijet dataset and (ii) its transformation by one of the generators discovered by the SymmetryGAN.
    Here, we plot the momenta of the two leading jets in the transverse plane.}
    \label{fig:LHCO_Comparison}
\end{figure*}

It is not easy to visualize a six-dimensional space, and the symmetries discovered by SymmetryGAN
do not lie in any single $2$-plane or even $3$-plane.
Therefore, we need alternative methods to verify that the maps discovered by the neural network are indeed symmetries.

One verification strategy is to visually inspect $X$ and $g_{\vb*\theta}(X)$ to see if the spectra look the same.
In \Fig{LHCO_Comparison}, we show a projection of the distribution of $X$ and one instance of $g_{\vb*\theta}(X)$, which suggests that the found $g_{\vb*\theta}$ is indeed a symmetry.

\begin{figure*}[t]
    \centering
    \subfloat[]{\includegraphics[width=0.45\textwidth]{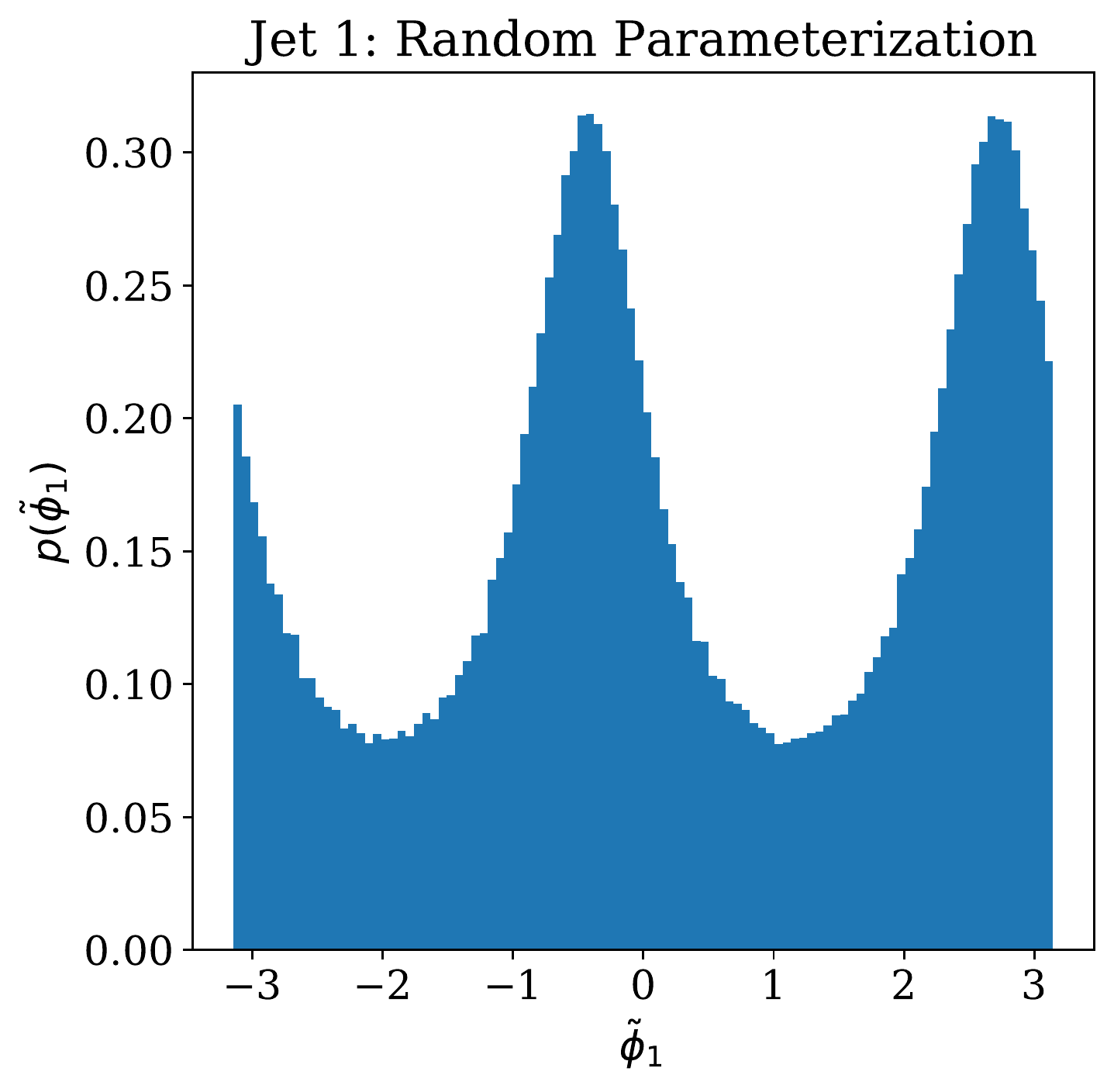} \label{fig:KLrand_i}}
    $\quad$
    \subfloat[]{\includegraphics[width=0.45\textwidth]{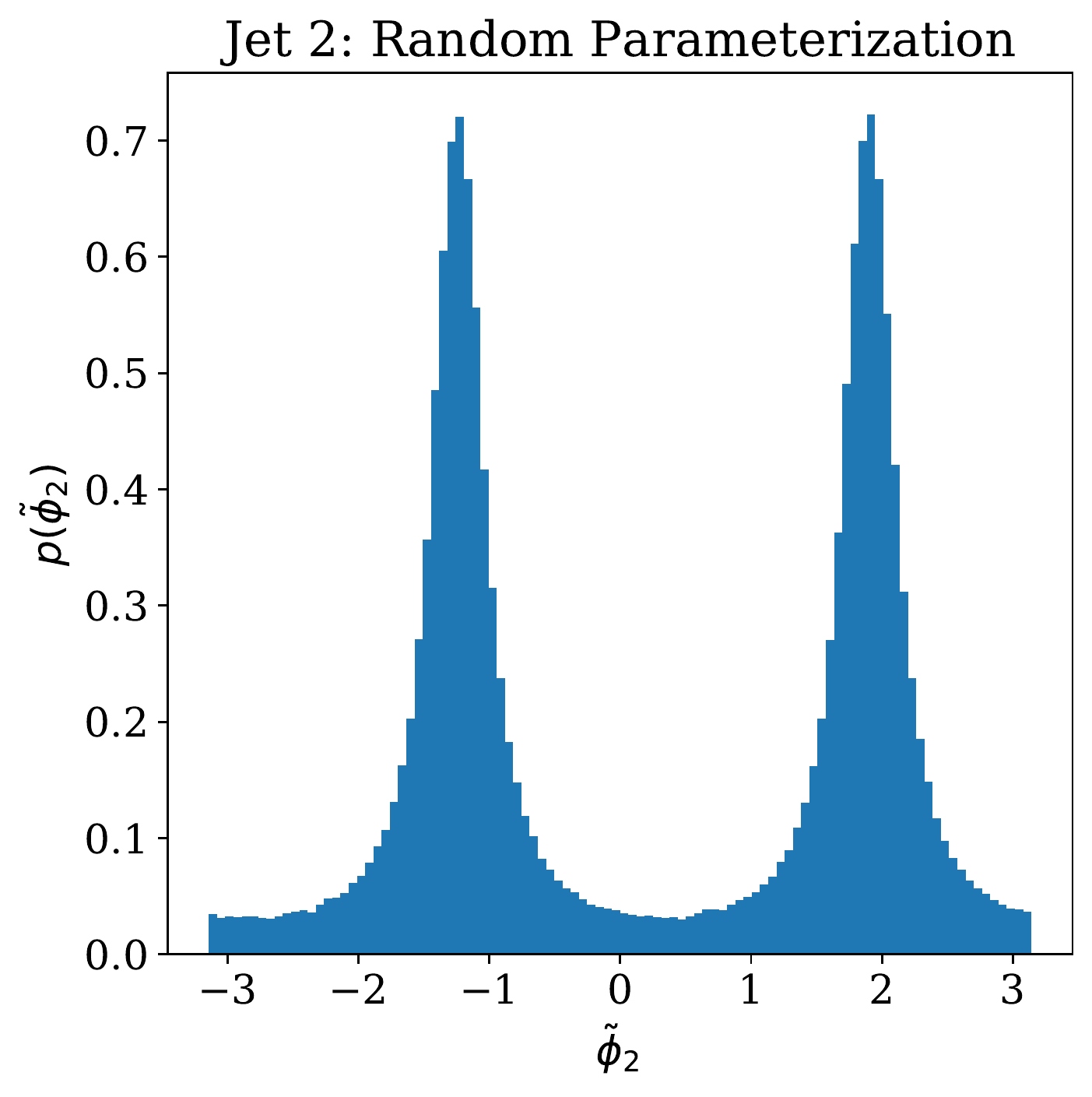}
    \label{fig:KLrand_ii}}
    \caption{
An example of the jet azimuthal angle distributions, (i)$\widetilde{\phi}_{1}$ and (ii)$\widetilde{\phi}_{2}$, of the LHC Olympics dijet data rotated by a randomly selected rotation in $SO(4)$.
The distribution is not uniform, so a random rotation is not a symmetry.}
    \label{fig:KL_rand}
\end{figure*}

\begin{figure*}[t]
    \centering
    \subfloat[]{\includegraphics[width=0.45\textwidth]{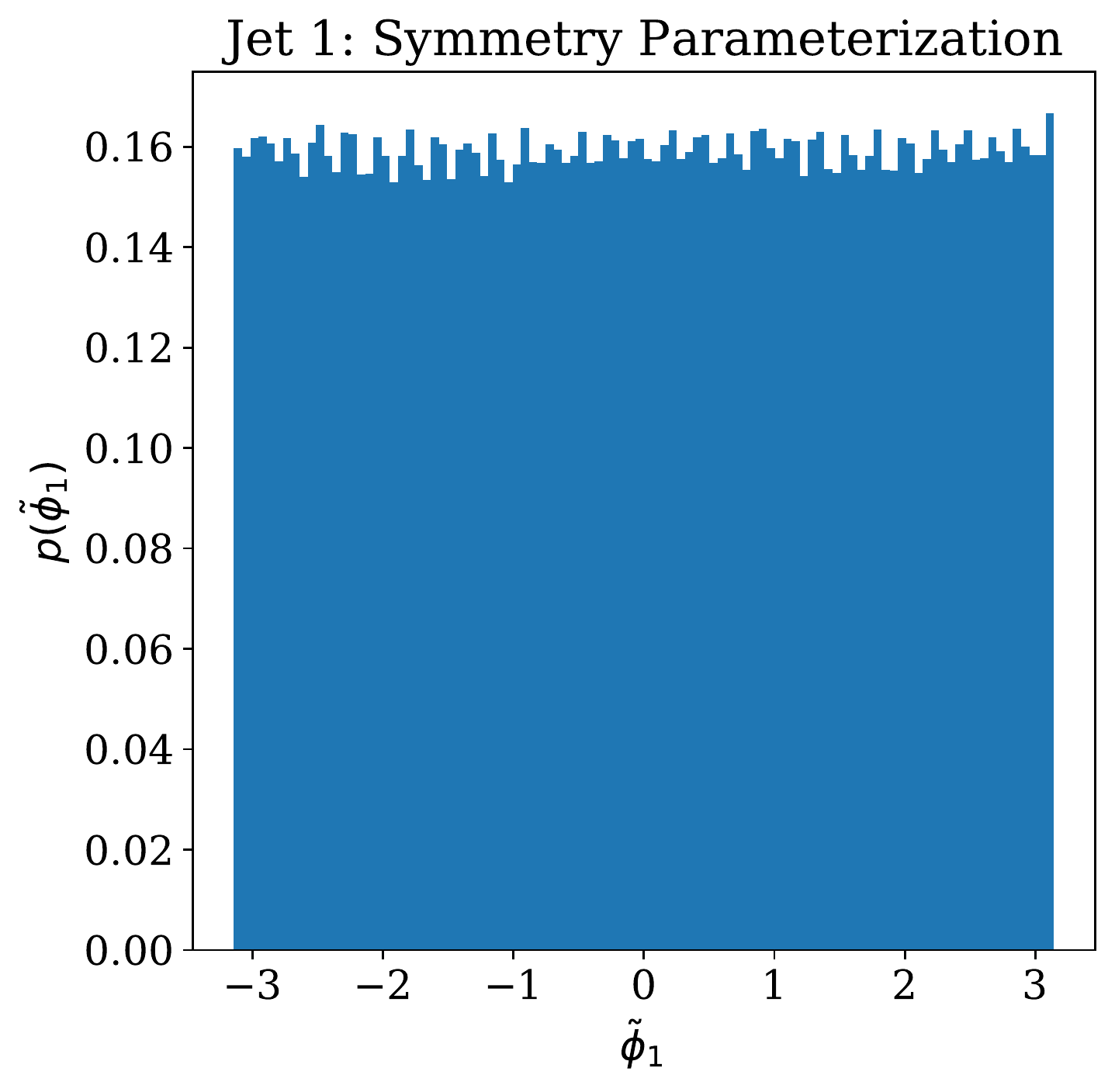} \label{fig:KLsymm_i}}
    $\quad$
    \subfloat[]{\includegraphics[width=0.45\textwidth]{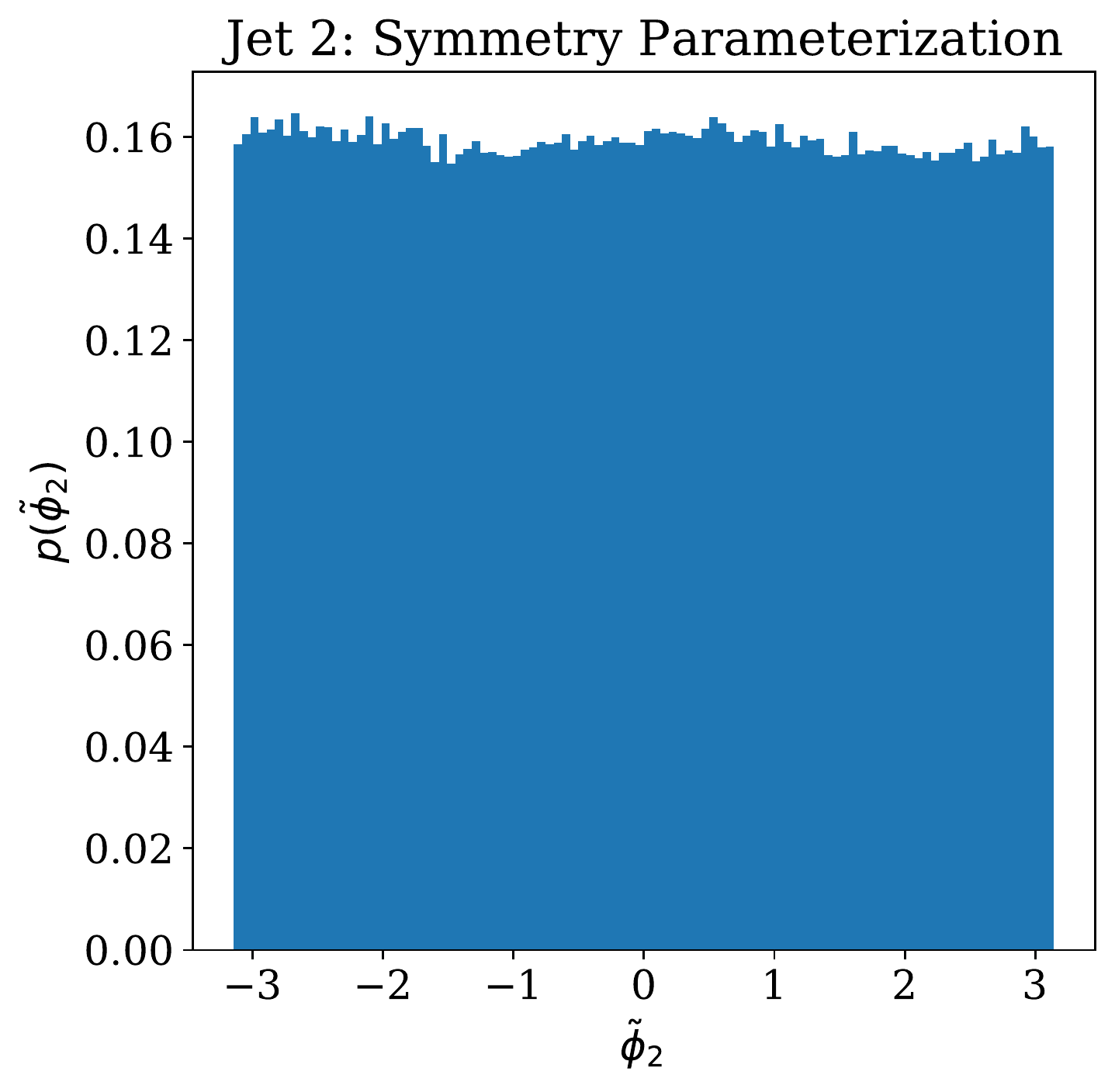}
    \label{fig:KLsymm_ii}}
    \caption{
    The same as \Fig{KL_rand}, but for a symmetry in $SO(4)$.
    The distribution is uniform, so this rotation is a candidate symmetry.
    }
    \label{fig:KL_symm}
\end{figure*}

Another verification strategy is to test if the discovered symmetries preserve special projections of the dataset.
Each of the two jets has an azimuthal angle $\phi_{j} = \text{arctan2}\,(p_{jy}, p_{jx})$ for $j = 1,2$ that is uniformly distributed over $[-\pi, \pi)$, where $\text{arctan2}$ is the two argument arctangent function, which returns the principal value of the polar angle $\theta\in (-\pi, \pi]$.
Symbolically, the data can be represented as
\begin{equation}
    X = \mqty[p_{1x}\\ p_{1y}\\p_{2x}\\p_{2y}] = \mqty[p_{1T}\cos\phi_{i1}\\p_{1T}\sin\phi_{1}\\p_{2T}\cos\phi_{2}\\p_{2T}\sin\phi_{2}],\qquad \phi_{j}\sim\U[-\pi, \pi)\,,
\end{equation}
where $p_{jT}$ is the transverse momentum of each jet (which is approximately the same for both jets since they are roughly back to back).
If one applies an arbitrary rotation, there is no reason the new azimuthal angles,
\begin{align}
    \widetilde{\phi}_{1} &= \text{arctan2}\,(g_{\vb*\theta}(X)_2, g_{\vb*\theta}(X)_1), \\
    \widetilde{\phi}_{2} &=  \text{arctan2}\,(g_{\vb*\theta}(X)_4, g_{\vb*\theta}(X)_3),
\end{align}
should be uniformly distributed anymore, as \Fig{KL_rand} demonstrates.
If one of the symmetry rotations discovered by the neural network is applied to $X$, however, $\widetilde{\phi}_{j}$ must remain uniformly distributed, as shown in \Fig{KL_symm}.

\begin{figure*}[t]
    \centering
    \subfloat[]{\includegraphics[width=0.45\textwidth]{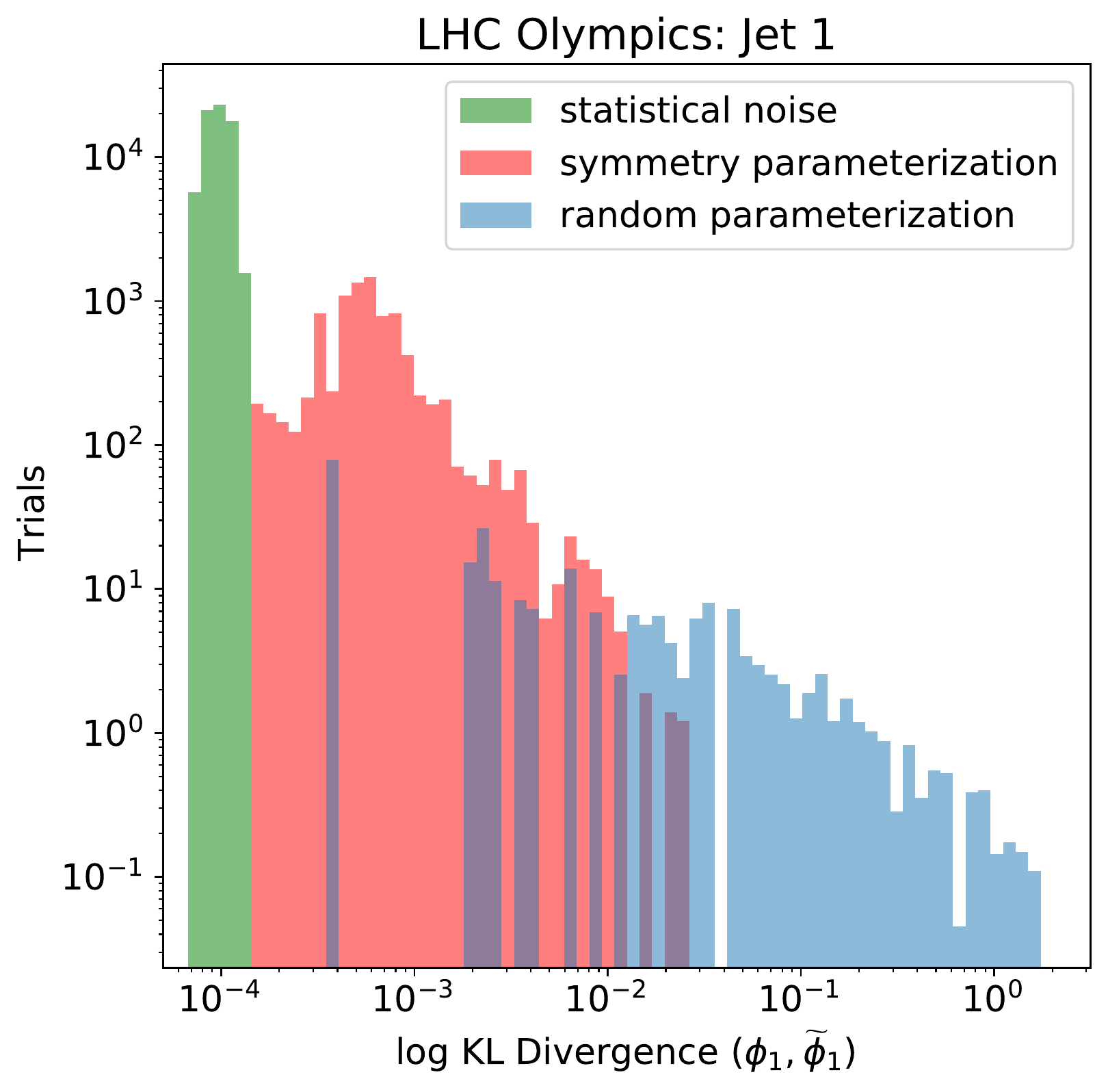} \label{fig:KLdiv_i}}
    $\quad$
    \subfloat[]{\includegraphics[width=0.45\textwidth]{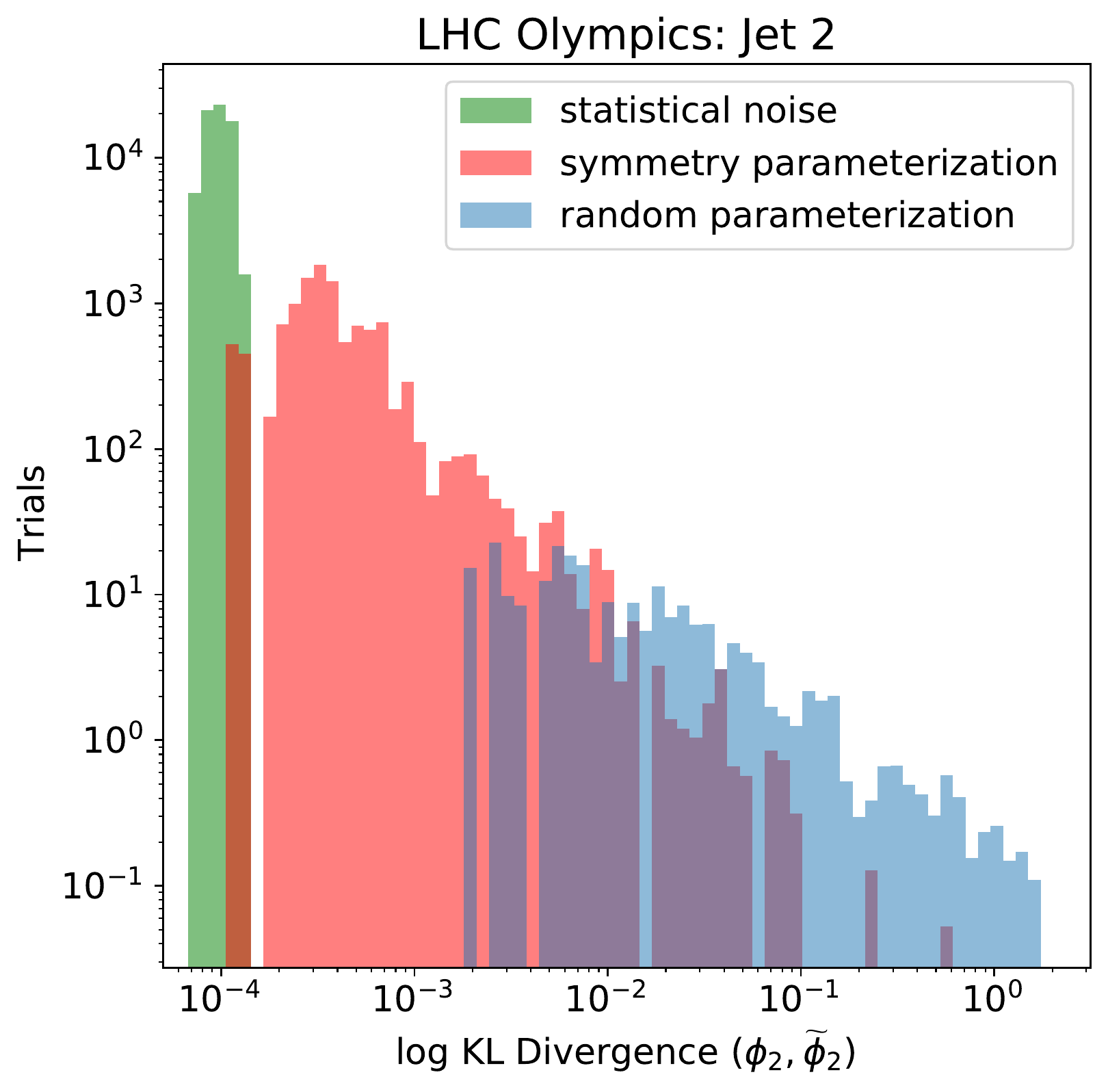}
    \label{fig:KLdiv_ii}}
    \caption{
    The KL divergence between the jet azimuthal angle distribution before and after a random rotation or a symmetry rotation, for the (i) leading jet and (ii) subleading jet.
    The KL divergence between two samples drawn from $\mathcal U[-\pi, \pi)$ is shown for comparison.
    }
    \label{fig:KL_div}
\end{figure*}

This effect can be quantified by computing the Kullback-Leibler (KL) divergence of the two $\widetilde{\phi}_{j}$ distributions against that of $\phi_{j}$.
In \Fig{KL_div}, we see that the KL divergence of the symmetries is much smaller than the KL divergence of the random rotations.
Also plotted on the same figure is the KL divergence of two samples drawn from $\U[-\pi, \pi)$, which represents the irreducible effect from considering a finite dataset.
This would be the KL divergence of $\widetilde{\phi}_{j}$ obtained from applying an ideal analytic symmetry to $X$, against $\phi_{j}$.
It is instructive to consider the means of the histograms.
The KL divergence of randomly selected elements of $SO(4)$ has means of $0.37$ ($0.34$) for the leading (subleading) jet, while the KL divergence of symmetries in $SO(4)$ has respective means $0.0058$ ($0.0090$).
The irreducible statistical noise has a mean of $0.0010$.

Clearly, the symmetries reconstruct the distribution much better than randomly selected elements of $SO(4)$, and are in fact quite close to the irreducible KL divergence due to finite sample size.
Note that the x-axis of \Fig{KL_div} is logarithmic, which magnifies the region near zero, so the difference between the symmetry histogram and the statistical noise histogram is smaller than it might appear.

\begin{figure}[t]
    \centering
    \includegraphics[width=0.4\textwidth]{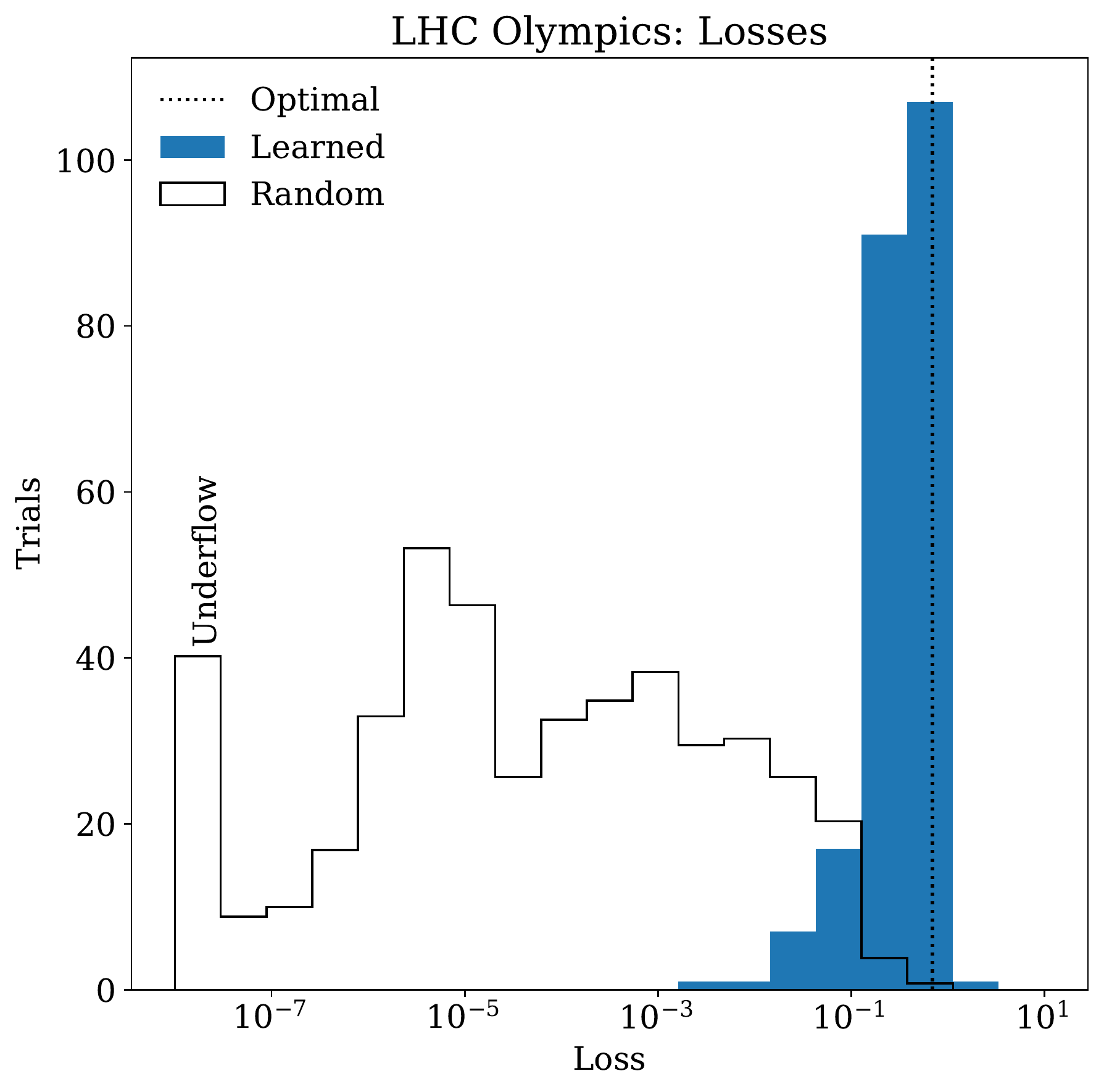}
    \caption{
    The loss of random rotations in $SO(4)$ compared to the loss of rotations learned by SymmetryGAN, overlaid with the analytic loss of a symmetry, $2\log2$.}
    \label{fig:LHCOLosses}
\end{figure}

A final method to independently verify that the rotations SymmetryGAN finds are symmetries of the LHC Olympics data is by computing the loss function.
As discussed at the end \Sec{1d_example}, when $g$ represents a symmetry and $d$ is an ideal discriminator, the binary cross entropy loss is $2\log2$.
By training a \textit{post hoc} classifier, we can therefore compute the loss of a specific symmetry generator.%
\footnote{In principle, one could look at the value of the loss after training the discriminator.  In practice, a post-hoc classifier yields more reliable behavior; see related discussion in \Ref{Diefenbacher:2020rna}.}
In \Fig{LHCOLosses}, we compare the loss of randomly sampled rotations from $SO(4)$ to the loss of rotations discovered by SymmetryGAN.
The latter is quite close to the analytic optimum, $2\log 2$.

From these tests, we conclude that SymmetryGAN has discovered symmetries of the LHC Olympics dataset.
As discussed further in \Sec{inference} below, though, discovering symmetries is different from inferring the structure of the found subgroup from the six-dimensional search space.
Mimicking the study from \Fig{MSE_i}, we can study its $\Z_2$ subgroups, through the loss function in \Eq{cyclicloss} with $q=2$.
The backbone of this subgroup is expected to be the reflections $p_{1k}\leftrightarrow p_{2k}$ (because both jets have approximately the same momenta) and $p_{jx}\leftrightarrow p_{jy}$ (because $\sin\phi$ and $\cos\phi$ look the same upon drawing sufficiently many samples of $\phi$).
The learning process reveals a much larger group, though.
There is in fact a continuous group of $\Z_2$ symmetries, which combine an overall azimuthal rotation and one of the aforementioned backbone reflections.
In retrospect, these $\Z_2$ symmetries should have been expected, since they are compositions of well-known symmetry transformations.
This example highlights the need to go beyond symmetry discovery and towards symmetry inference.

\section{Towards Symmetry Inference}
\label{sec:inference}

The examples in \Secs{results}{hepexample} highlight the potential power of SymmetryGAN for discovering symmetries using deep learning.
Despite the many maps discovered by the neural network, though, it is difficult to infer, for example, the precise Lie subgroup of $SO(4)$ respected by the LHC Olympics data.
This highlights a limitation of this  approach and the distinction between ``symmetry discovery'' and ``symmetry inference''.
Though SymmetryGAN can identify points on the Lie group manifold, there is no simple way to infer precisely which Lie group has been discovered.
While symmetry discovery is sufficient for the data augmentation described in previous sections to facilitate data analysis, it is of theoretical interest to infer which formal Lie groups comprise the symmetries of our collider data.
In this section, we mention three potential methods to assist in the process of symmetry inference.

\subsection{Finding Discrete Subgroups}
\label{sec:discrete_subgroups}

One way to better understand the structure of the learned symmetries is to look for discrete subgroups.
As already shown in \Fig{MSE} and mentioned in the particle physics case, we can identify discrete $\Z_q$ symmetry transformations by augmenting the loss with \Eq{cyclicloss}.
By forcing the symmetries to take a particular form, we can infer the presence (or absence) of such a subgroup.

It is interesting to consider possible modifications to \Eq{cyclicloss} to handle non-Abelian discrete symmetries.
The goal would be to learn multiple symmetries simultaneously that satisfy known group theoretic relations.
For example in the Abelian case, a loss term like 
\begin{equation}
\label{eq:abeliansymm}
    -\frac\alpha N\sum_{x\in\{x_i\}_{i=1}^N}(g_1 \circ g_2(x) - g_2 \circ g_1(x))^2
\end{equation}
could be used to identify any two symmetries $g_1$ and $g_2$ that commute.
We leave a study of these possibilities to future work.

\subsection{Group Composition}

By running SymmetryGAN a few times, one may discover a few points on the symmetry manifold.
By composing these discovered symmetries together, one can rapidly increase the number of known points on the manifold because the discovered symmetries are elements of a group, by construction, so their composition is still an element of the group.

This notion is quite powerful.
The ergodicity of the orbits of group elements is a richly studied and complex area of mathematics (see e.g.~\Ref{bams/1183548783}).
Many groups of physical interest are locally connected, compact, and have additional structure.
In that context, it is likely that the full symmetry group is generated by $\qty{r_1,\dots, r_\nu}$, where $r_i$ is randomly drawn from the group and $\nu$ is the product of the representation dimension and the number of connected components.

For example, consider the group $U(1)\cong SO(2)$, which has $\nu=1$.
Almost any element of $U(1), e^{i\theta}$, has rotation angle which is an irrational multiple of $\pi, \frac\theta\pi\in\R\setminus\mathbb{Q}$.
We can therefore approximate any element $e^{i\phi}\in U(1)$ by repeated applications of $e^{i\theta}$:
\begin{equation}
    \forall e^{i\phi}\in U(1)\;\forall\epsilon > 0\;\exists n\in \mathbb{N}\; \norm{e^{i\phi} - e^{in\theta}} < \epsilon\,.
\end{equation}
In other words, the subgroup generated by $e^{i\theta}$ is dense in $U(1)$.

In practice, the symmetries discovered by SymmetryGAN will be not exact due to numerical considerations.
Since the network learns approximate symmetries with some associated error, each composition compounds this error.
Thus, there are practical limits on the number of compositions that can be carried out with numeric data.

\subsection{The Symmetry Discovery Map}
\label{sec:symmetry_discovery_map}

So far, we have initialized a SymmetryGAN with uniformly distributed values of certain parameters, and then trained it to return the values of those parameters that constitute a symmetry.
We can define a \textit{symmetry discovery map}, which connects the initialized parameters of $g$ to the parameters of the learned function:
\begin{equation}
\Omega: \R^k \to \R^k,
\end{equation}
where $k$ is the dimension of the parameter space.
This is a powerful object not only for characterizing the learning dynamics but also to assist in the process of symmetry discovery and inference.

There are at least two distinct reasons why knowledge of this symmetry discovery map is useful.
First, the map is of theoretical interest.
We discussed in \Sec{1d_example} the importance of understanding the topology of the symmetry group. 
The symmetry discovery map induces a \emph{deformation retract} from the search space to the symmetry space.
Every deformation retract is a homotopy equivalence, and by the Eilenberg-Steenrod axiom of homotopy equivalence~\cite{Eilenberg117}, the homology groups of the symmetry group can be constructed from the homology groups of the search space.
Even in low dimensions, the topology of the symmetry group can be non-trivial (cf.~\Sec{2d_example} for an example in 2D).
The topology of $GL_n(\R)$, however, has been studied for over half a century, and the homotopy and homology groups of several non-trivial subgroups of $\operatorname{Aff}_n(\R)$ have been fully determined~\cite{SCHLICHTING20171}.
Hence, if the symmetry discovery map were known, one could leverage the full scope of algebraic topology and the known results for the linear groups to understand the topology of the symmetry group.

Second, this map has practical value.
Every time a SymmetryGAN is trained, it must relearn how to move the initialized values of $g$ to the final values.
Intuitively, nearby initial values should map to nearby final values, so learning the symmetry discovery map should enable a more efficient exploration of the symmetry group.
In practice, this can be accomplished by augmenting the loss function in \Eq{numericloss}.
Let $g(x|c)$ be the symmetry generator, with the parameters $c$ made explicit.
Let $\Omega(c)$ be a neural network representing the symmetry discovery map.
Sampling parameters from the space of parameters $\R^k$ and data points from $X$, we can optimize the following loss:
\begin{align}
\label{eq:numericloss_parameters}
    L[\Omega,d]&=-\sum_{c \in \{c_a\}} \sum_{x\in\{x_i\}}  \Big[\log\big(d(x)\big) \\
    & \nonumber \qquad \qquad \qquad \qquad + \log\big(1-d(g(x|\Omega(c)))\big)\Big]\,.
\end{align}
Note that this loss is now a functional of $\Omega$ instead of $g$.
If $\Omega(c)$ can be initialized to the identity function, then gradient descent acting on $\Omega(c)$ is (asymptotically) the same as gradient descent acting on the original parameters.
Thus, as long as $\Omega(c)$ has a sufficiently flexible parametrization, the learned $\Omega(c)$ will be a good approximation to the symmetry discovery map learned by the original SymmetryGAN.

We defer a full exploration of the symmetry discovery map to future work.
Preliminary analytic and numerical studies of the symmetry discovery map are shown in \App{symmetry_discovery_map}.

\section{Conclusions and Outlook}
\label{sec:conclusions}

In this paper, we provided a rigorous statistical definition of the term ``symmetry'' in the context of probability densities.
This is highly relevant for the field of high energy collider physics where the key objects of study are scattering cross sections.
We proposed SymmetryGAN as a novel, flexible, and fully differentiable deep learning approach to symmetry discovery.
SymmetryGAN showed promising results when applied to Gaussian datasets as well as to dijet events from the LHC, conforming with our analytic predictions and providing new insights in some cases.

A key takeaway lesson is that the symmetry of a probability density only makes sense when compared to an inertial density.
For our studies, we focused exclusively on the inertial density corresponding to the uniform distribution on (an open subset of) $\R^n$, since Euclidean symmetries are ubiquitous in physics.
Furthermore, we only considered area preserving linear maps on $\R^n$, a simple yet rich group of symmetries that maintain this inertial density.
This method has great utility for data analysis.
The symmetries of a dataset discovered by SymmetryGAN can be used to augment a dataset, thereby increasing its statistical power substantially.
Conversely, it could be used to preprocess the data to explicitly project out symmetries and fix a preferred reference frame, thereby once again boosting the data analysis process substantially.
Moving forward, there are many opportunities to further develop the concepts introduced in this paper.
As a straightforward extension, non-linear equiareal maps over $\R^n$ could be added to the linear parametrizations we explored, as could Lorentz-like symmetries.
In more complex cases where there is no obvious notion of an inertial density, one could study the relative symmetries between two different datasets.
It would also be interesting to discover approximate symmetries and rigorously quantify the degree of symmetry breaking.
This is relevant in cases where the complete symmetry group is obscured by experimental acceptances and efficiencies.

A key open question is how to go beyond symmetry discovery and towards symmetry inference.
We showed how one can introduce loss function modifications to emphasize the discovery of discrete subgroups.
One could imagine extending this strategy to continuous subgroups to gain a better handle on group theoretic structures.
The symmetry discovery map is a potentially powerful tool for symmetry inference, since it in principle allows the entire symmetry group to be discovered in a single training.
In practice, though, we found learning the symmetry discovery map to be particularly challenging.
We hope future algorithmic and implementation developments will enable more effective strategies for symmetry discovery and inference, in particle physics and beyond.

The code for this paper can be found in this \href{https://github.com/hep-lbdl/symmetrydiscovery}{\underline{GitHub repository}}.

\section*{Acknowledgments}

We would like to thank Andrew Larkoski for his thoughtful comments and feedback.
KD and BN are supported by the U.S. Department of Energy, Office of Science under contract DE-AC02-05CH11231.
JT is supported by the National Science Foundation under Cooperative Agreement PHY-2019786 (The NSF AI Institute for Artificial Intelligence and Fundamental Interactions, \url{http://iaifi.org/}), and by the U.S. DOE Office of High Energy Physics under grant number DE-SC0012567.

\appendix

\begin{figure*}
    \centering
    \subfloat[]{
    \includegraphics[width=0.5\textwidth]{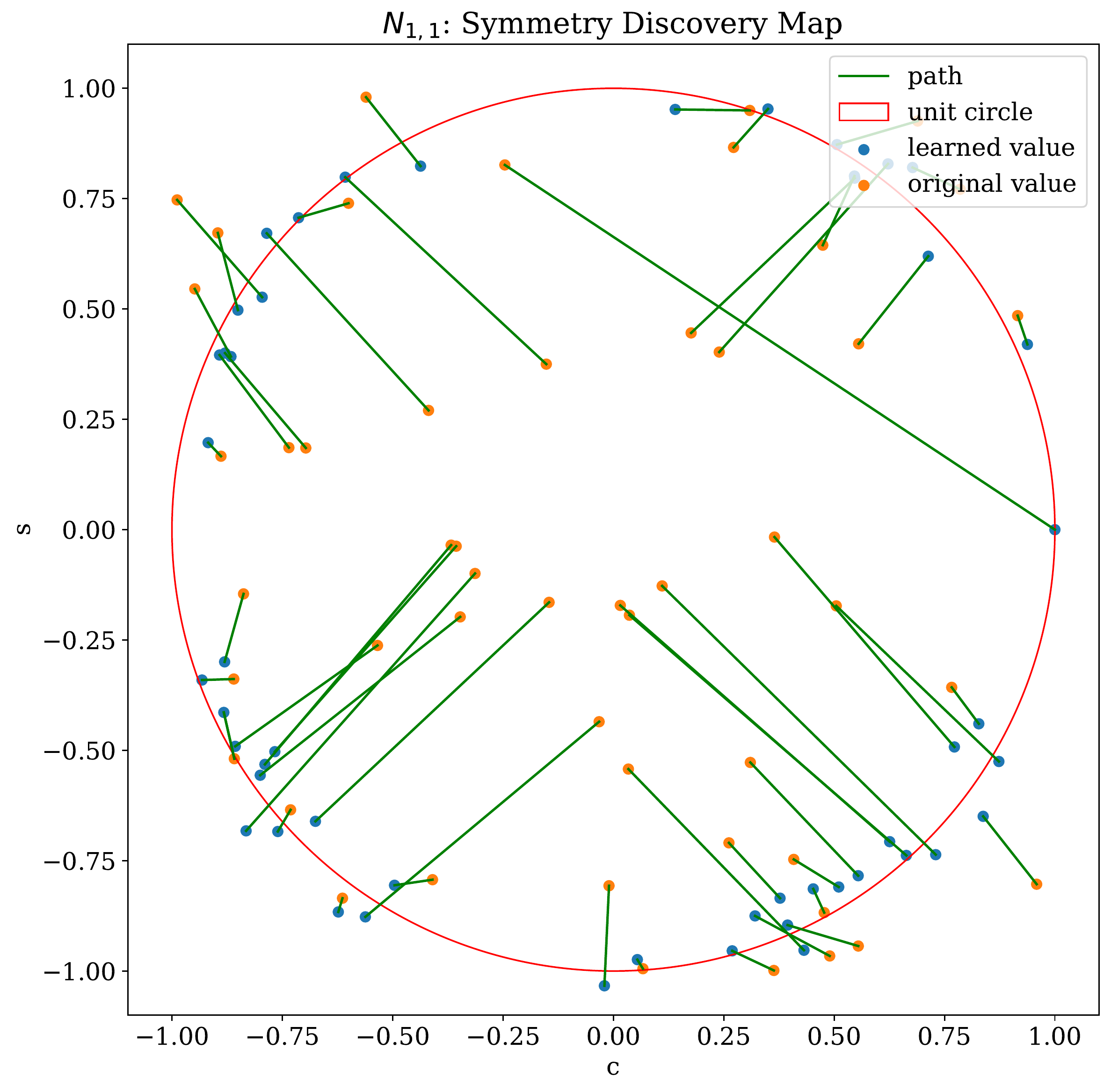}
       \label{fig:SO2symmi2f}
       }
    \subfloat[]{
       \includegraphics[width=0.5\textwidth]{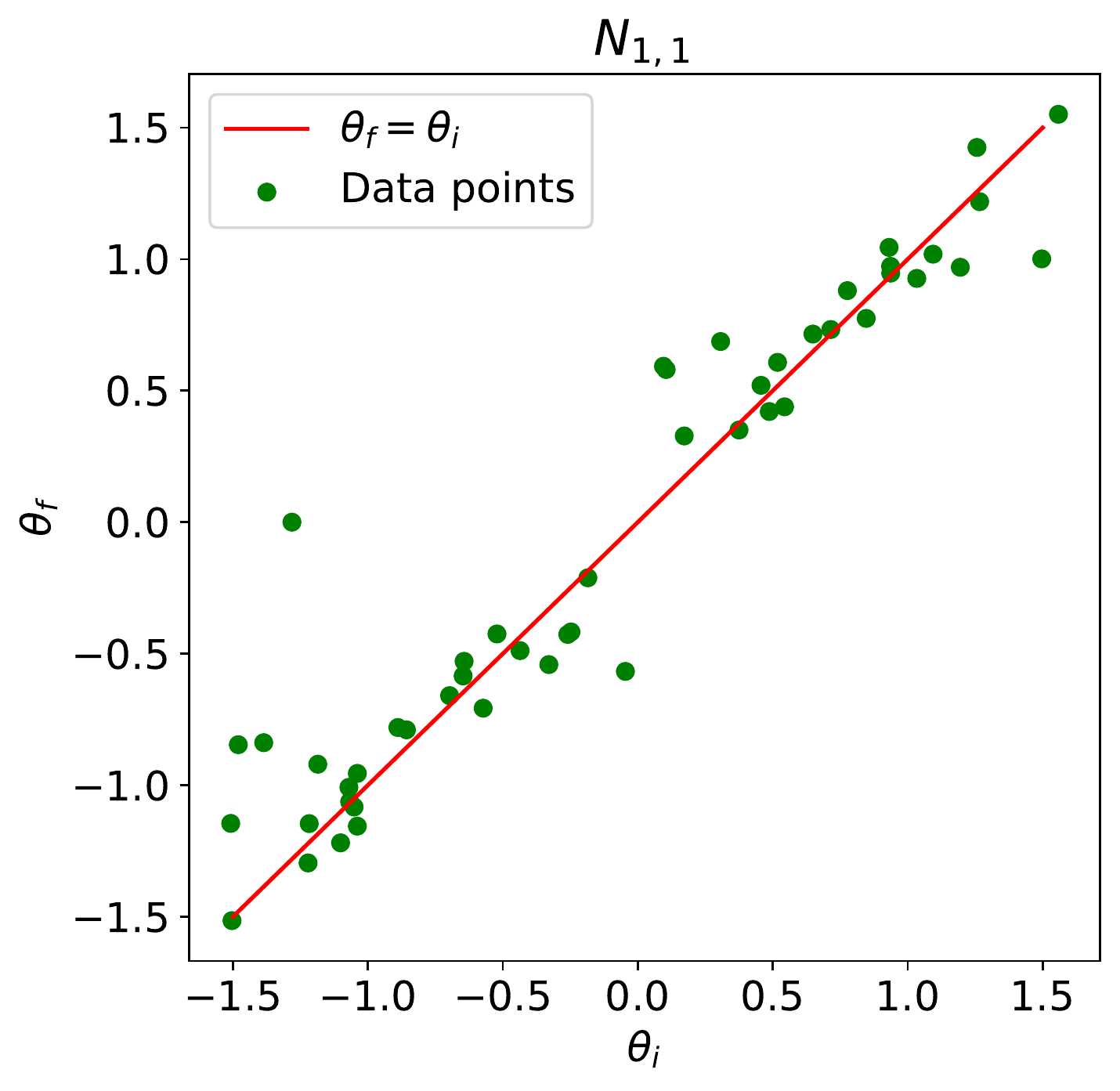}
       \label{fig:SO2symmtheta}
       }
    
    \caption{Symmetry discovery maps for the standard two-dimensional Gaussian.
    (i) Motion from the initialized parameters to the learned parameters.
    (ii)
    Transforming to polar coordination, the initialized and learned values of the polar angle.}
\end{figure*}

\section{Explorations of the Symmetry Discovery Map}
\label{app:symmetry_discovery_map}

In this appendix, we initiate a study of the symmetry discovery map from \Sec{symmetry_discovery_map}, both analytically and numerically.

\subsection{One-Dimensional Gaussian}

In the simplest cases, the symmetry discovery map can be determined analytically, and a neural network can be used to independently verify that the proposed map is indeed the symmetry discovery map.
For example, consider the one-dimensional Gaussian example from \Sec{1d_example}, where the probability distribution is $\mathcal N(0.5, 1)$ and the candidate symmetry transformations take the form $g(x) = b + cx$ for $(b, c)\in\R^2$.
There are two symmetries in this case: the identity and $g(x)=1-x$.

In \Sec{1d_example}, we conjectured that the learned symmetry is the one on the same side of the loss barrier at $c = 0$ as the initialization.
This means that independent of $b_i$, if $c_i >0$, then $g$ will be the identity and if $c_i < 0$, $g$ will be the inversion map.
Symbolically, the symmetry discovery map $\Omega:\R^2\to\R^2, \Omega: (b_i, c_i)\mapsto (b_f, c_f)$ takes the form
\begin{align}
\Omega(b, c) = \begin{cases}(0,\;\;1)& c > 0 \, ,\\ (1, -1) & c < 0 \,.
\end{cases}
\end{align}
The numerical results already shown in \Fig{Z2numeric} verify that $\Omega$ is indeed the correct symmetry discovery map.

\subsection{Two-Dimensional Gaussian}

We next consider one of the two-dimensional Gaussian examples from \Sec{2d_example}.
The probability distribution is $N_{1,1}$ and the candidate symmetry transformations are 
\begin{equation}
    g(X) = \mqty[c&s\\-s&c]X, \qquad (c, s)\in\R^2.
\end{equation}
From analyzing the loss landscape, we expect the neural network to map the initialized point to the nearest point on $SO(2)$ along a radius of the unit circle.
This leads to the symmetry discovery map:
\begin{align}
\Omega(c, s) = \qty( \frac{c}{\sqrt{c^2 + s^2}}, \frac{s}{\sqrt{c^2 + s^2}})\,.
\end{align}
In \Fig{SO2symmi2f}, we show the numerical mapping between initial and final parameters, corresponding to the plot in \Fig{SO2_i}.
The radial behavior is clearly visible, although there are some outliers that could be due to incomplete training and the stochastic nature of the gradient decent.

We can gain more insight by studying this behavior in polar coordinates: 
\begin{align}
r = \sqrt{c^2 + s^2}\qquad \theta = \text{arctan2}\,(s, c)\,.
\end{align}
Going back to $c$ and $s$ can be done with the inverse mapping $c = r\cos\theta$ and $s = r\sin\theta$.
In polar coordinates, the symmetry discovery map is rather simple:
\begin{equation}
\Omega(r, \theta) = (1, \theta).
\end{equation}
The numerics support this prediction, as shown in \Fig{SO2symmtheta}.
The initialized and learned points collect around the line of constant polar angle.

\subsection{Learning the Symmetry Discovery Map}
\label{app:numerics}

Ultimately, the symmetry discovery map will be most useful if it can be learned from a single training run.
In preliminary studies, however, we encountered two key challenges.

The first challenge is that, for $\Omega$ to approximate the symmetry discovery map, it needs to be initialized to the identity function.
If the goal is just to find a family of symmetries, then it would be fine to start from a randomly initialized neural network.
In that case, $g(x|\Omega(c))$ would be a parametrized symmetry network, in the spirit of \Ref{Baldi:2016fzo}.
But for the goal of finding the symmetry discovery map, one needs a parametrization of $\Omega$ that is flexible enough to describe the map, but simple enough that it can be initialized close to the identity.

The second challenge is that performing min-max optimization of \Eq{numericloss_parameters} seems particularly finicky.
GANs are known to exhibit issues like mode collapse, and because the target space of the symmetry discovery map is often disjoint, these kinds of issues seem to arise in our case as well.

Consider the simple case of learning the symmetry discovery map for data $X\sim\mathcal{N}(0, 1)$ and the generator $g(x) = c\, x$. 
We know that the symmetries of $X$ are $g(x) = \pm x$.
Therefore, the symmetry discovery map should be the step function $\Omega(c) = \operatorname{sign}(c)$:
\begin{equation}
    \label{eq:step_discovery}
    \Omega(c) = \begin{cases}1 & c > 0\, .\\
    -1 & c < 0\,.
    \end{cases}
\end{equation}
The form in \Eq{step_discovery} is not so easy to learn with any of the standard neural network activation functions, though.
The one exception is unit step activation, of course, but this activation is far from the identity and therefore difficult to use for finding a symmetry discovery map.

One approach to this problem is to use a custom activation function:
\begin{equation}
    \Omega(c) = \lambda\,\relu(c) - \mu\,\relu(-c) + \rho\, .
\end{equation} 
This can be initialized at $\lambda = \mu = 1$, $\rho = 0$ so that in the beginning $\Omega = \mathbbm{1}$.
As $\lambda$ moves away from $\mu$, this function develops a non-linearity as desired.
With a single layer, this form is not sufficient to learn the correct answer, though it may be possible that with a deep network stacked with these components, the correct map could be learned.

Another approach to this problem is to use polynomial activation,
\begin{equation}
    \Omega(c) = \lambda c + \mu c^2 + \nu c^3 + \dots + \zeta c^{11},
\end{equation}
initialized with $\lambda = 1, \mu = \nu = \dots = \zeta = 0$.
A step function is not within this class of functions, but with such a high degree polynomial, it is expected to be a reasonable approximation.
This was in fact the case, though the result was far from satisfactory.

Finally, as a proof of principle, we tested the simplified ansatz:
\begin{equation}
    \Omega(c) = \lambda c + (1-\lambda) \operatorname{sign} (c - \mu).
\end{equation}
When initialized with the identity function ($\lambda = 1, \mu = 0$), gradient descent indeed converges to \Eq{step_discovery} ($\lambda = 0, \mu = 0$).
This ansatz is too contrived to draw any robust conclusions, but is does motivate future exploration of more complex architectures and training protocols to learn the symmetry discovery map.

\bibliography{HEPML,other}

\end{document}